\documentclass[aps,superscriptaddress,eqsecnum,tightenlines,preprint,nofootinbib]{revtex4}
\usepackage{graphicx,amssymb,amsbsy,bm,amsmath}

\usepackage{hyperref}

\newcommand{\Ok}{\ensuremath{\Omega_k}}

\newcommand{\vx}{\ensuremath{\vec{x}}}

\newcommand{\vy}{\ensuremath{\vec{y}}}
\newcommand{\vk}{\ensuremath{\vec{k}}}
\newcommand{\vq}{\ensuremath{\vec{q}}}
\newcommand{\vp}{\ensuremath{\vec{p}}}

\newcommand{\be}{\begin{equation}}
\newcommand{\ee}{\end{equation}}
\newcommand{\bea}{\begin{eqnarray}}
\newcommand{\eea}{\end{eqnarray}}

\begin{document}
\title{An effective field theory during inflation II: \\ stochastic dynamics and power spectrum suppression.}
\author{D. Boyanovsky}
\email{boyan@pitt.edu} \affiliation{Department
of Physics and Astronomy,\\ University of Pittsburgh\\Pittsburgh,
Pennsylvania 15260, USA}
\date{\today}
\begin{abstract}
We obtain the non-equilibrium effective action of an inflaton like scalar field --the system-- by tracing over sub Hubble degrees of freedom  of ``environmental'' light scalar fields. The effective action is  stochastic leading to   effective Langevin equations of motion for the fluctuations of the inflaton-like field,  with self-energy corrections and stochastic noise correlators that obey a de Sitter space-time analog of  a fluctuation dissipation relation.   We solve the Langevin equation  implementing a dynamical renormalization group resummation of the leading secular terms and obtain the corrections to the power spectrum of super Hubble fluctuations of the inflaton field, $\mathcal{P}(k;\eta) = \mathcal{P}_0(k)\,e^{-\gamma(k;\eta)}$ where $\mathcal{P}_0(k)$ is the nearly scale invariant power spectrum in absence of coupling. $\gamma(k;\eta)>0$ describes the suppression of the power spectrum, it features Sudakov-type double logarithms and entails violations of scale invariance.  We also obtain the effective action for the case of a heavy scalar field of mass $ M \gg H$, this case yields a local ``Fermi'' limit with a very weak self-interaction of the inflaton-like field and dissipative terms that are suppressed by powers of $H/M$. We conjecture on the possibility that the large scale anomalies in the CMB may originate in dissipative processes from inflaton   coupling to sub-Hubble degrees of freedom.

\end{abstract}


\maketitle

\section{Introduction}\label{sec:intro}
Precise observations of the cosmic microwave background anisotropies by the WMAP\cite{wmap} and PLANCK\cite{planck} missions support one of the main predictions of inflationary cosmology, namely a  nearly scale invariant power spectrum of adiabatic perturbations. A main paradigm of inflationary cosmology posits that the inflationary stage is dominated by a   scalar field slowly rolling down a potential landscape leading to a nearly de Sitter inflationary stage. Cosmological perturbations   are generated by  quantum fluctuations that are amplified when their wavelengths become larger than the Hubble radius during inflation.   Most models of inflation rely on either a single scalar field or several scalar fields (which typically yield isocurvature  or entropy  perturbations), however the effect of the interactions between the inflaton field and the degrees of freedom  that would populate a radiation dominated era post inflation is neglected during the inflationary stage and \emph{assumed} to switch-on at the end of inflation during a period of ``reheating''\cite{reheat1,reheatrev}. However, such   interaction between the inflaton and other (fermionic,   scalar) degrees of freedom cannot just switch-on at the end of inflation, on physical grounds  if the inflaton couples to other degrees of freedom, such couplings should also influence the inflationary stage.

Interactions of quantum fields in de Sitter (or nearly de Sitter) space-time have been the focus of important studies\cite{woodardcosmo,proko1,decayds,akhmedov,woodard1,prokowood,onemli,sloth1,sloth2,riotto,fermionswoodpro,picon,lello,rich,raja,
richboy,boyan,serreau,parentani,smit,polyakov} which show strong infrared and secular effects.

 Furthermore, non-Gaussianity is a manifestation of self-interactions of curvature perturbations and could leave an observable imprint on the cosmic microwave background, although it is    suppressed by small slow roll parameters  in    single field slow roll inflationary models\cite{maldacena,komatsu}.

Interactions with heavy fields with masses $M\gg H$ with $H$ the Hubble parameter during inflation have been treated in terms of effective field theory descriptions\cite{heavyachu,heavycespe,heavy1,heavy2} mainly by neglecting kinetic terms and correlations,   treating the heavy degrees of freedom as auxiliary fields that can be ``integrated out'' at tree level, or including correlations of the heavy fields in powers of $H/M \ll 1$\cite{heavyjack}.

In a non-equilibrium situation as is the case with cosmological expansion,   integrating out   short wavelength degrees of freedom leads to an effective field theory description in terms of a \emph{reduced density matrix} for long wavelength fluctuations. Such a description is, fundamentally, akin to a \emph{Wilsonian} approach to an effective field theory\cite{bala} by coarse graining short distance degrees of freedom.  At the level of a non-equilibrium effective action, the study of the effects of tracing out degrees of freedom  was pioneered with the study of quantum Brownian motion\cite{feyn,schwinger,caldeira,calzetta,paz,hu,leticia}, the degrees of freedom of interest are considered to be the ``system'' whereas those that are integrated out (traced over) are the ``bath'' or ``environment''. The effects of the bath or environment are manifest in the non-equilibrium effective action via an \emph{influence action} which is in general non-local and describes dissipative processes. This influence action is   determined by the \emph{correlation functions} of the environmental degrees of freedom,  and determines the time evolution of the reduced density matrix. An alternative   description of the time evolution of the reduced density matrix is the \emph{quantum master equation}\cite{breuer,zoeller} which includes the effects of coupling to the environmental degrees of freedom via their quantum mechanical correlations. In ref. \cite{boyeff} the equivalence between the influence action  and the  quantum master equation    was established in Minkowski space-time,  and shown that they provide a non-perturbative resummation of self-energy diagrams directly in real time providing an effective  field theory description of non-equilibrium phenomena.

 A generic quantum master equation approach   for a reduced density matrix describing cosmological perturbations has been advocated in ref.\cite{burhol} in terms of local correlations of environmental degrees of freedom. More recently   the quantum master equation that describes the non-equilibrium dynamics of an inflaton-like scalar field fluctuations upon tracing out ``environmental'' sub-Hubble degrees of freedom was obtained\cite{boydensmat}. This quantum master equation yields the effective equations of motion for fluctuations of the inflaton field whose solution represents a non-perturbative resummation of self-energy diagrams. The results of this study reveals a suppression of the power spectrum of super Hubble fluctuations as a consequence of the interaction with the ``environmental'' degrees of freedom.

\vspace{2mm}

\textbf{Motivations and objectives:}
Although the inflationary paradigm has been successfully supported by precise observations of the power spectrum of temperature anisotropies in the cosmic microwave background (CMB), there remain puzzling anomalies at large scales\cite{starkman}, such as the suppression of power at low l's.  The preliminary results of ref.\cite{boydensmat} suggest that the coupling of the inflaton field to sub-Hubble degrees of freedom \emph{may} lead to dissipative effects that yield a suppression of the power spectrum on large scales. The confirmation of this suppression with the complementary framework of the non-equilibrium effective action, including the \emph{influence action} from integrating out sub-Hubble degrees of freedom is one of main  the goals of this article.

The equivalence between the influence action and the quantum master equation established in ref.\cite{boyeff} relies on a spectral representation of the correlation functions of the bath which is not available in inflationary cosmology as a consequence of the explicit time dependence of the gravitational background. In this article we obtain the non-equilibrium effective action including the \emph{influence action} that emerges from integrating out the ``environmental'' fields. This   approach  leads directly to a stochastic description in terms of an effective Langevin equation of motion\cite{feyn,calzetta,paz,hu,boyeff,serlan} it is complementary, and here shown to be equivalent to, the density matrix formulation of ref.\cite{boydensmat}.
 We seek to confirm and expand the results   obtained in  reference \cite{boydensmat} directly from the non-equilibrium effective action with the influence action, with the added bonus of making explicit the stochastic nature of the non-equilibrium effective action.   In the quantum density matrix this stochastic nature of the dynamics is not manifest directly and requires an interpretation in terms of Wiener processes\cite{breuer,zoeller}, whereas the influence action approach leads directly to a stochastic description of the non-equilibrium dynamics.  This is an important feature of the non-equilibrium effective action.
The pioneering work of ref.\cite{staro} recognized that integrating out sub-Hubble components of the inflaton scalar field during inflation yields a stochastic effective description. Several studies showed that decoherence and effective stochastic dynamics emerging from tracing over short wavelength degrees of freedom  are   of fundamental importance in cosmology\cite{staro,calhuuniv,stoca,proko1,woodard1,prokowood,lee,garbrecht}.
In this article we have two main objectives: i) To obtain the effective field theory that describes the non-equilibrium time evolution of the reduced density matrix for the ``system degrees of freedom'' by tracing out environmental degrees of freedom obtaining the influence action. This approach yields a stochastic description and complements and extends the study in ref.\cite{boydensmat}, with the goal of confirming the large scale suppression of the power spectrum as a consequence of dissipative effects.   ii) To  study the emergence of a local ``Fermi'' effective field theory in the case when the environmental field features a mass $M\gg H$, by obtaining the non-equilibrium effective action in powers of $H/M$ and establishing the nature of the dissipative contributions to the effective action in this case.

\vspace{2mm}

\textbf{Brief summary of results:} We obtain an effective field theory for an inflaton-like scalar field by tracing out (integrating) out another scalar field --the environment-- from the time evolved quantum density matrix. The effect of the environmental degrees of freedom is encoded in an \emph{influence action}  that leads to a stochastic description of the dynamics in terms of a Langevin equation that includes self-energy corrections and a colored noise. The self-energy and the noise correlation function obey a de Sitter space-time analog of the fluctuation-dissipation relation. The environmental fields are taken to be  massless and  conformally coupled to gravity as a \emph{proxy} for degrees of freedom that remain sub-Hubble during inflation, but we also analyze the case of a nearly massless minimally coupled field.   We implement the dynamical renormalization group to solve the Langevin equation and obtain the power spectrum of super Hubble inflaton fluctuations. The power spectrum is \emph{suppressed} as a consequence of the dissipative processes arising from the coupling to the environmental fields. For a nearly massless inflaton field for which the unperturbed power spectrum is scale invariant, we find
\be \mathcal{P}(k;\eta) = \mathcal{P}_0(k)\,e^{-\gamma(k,\eta)}~~;~~\gamma(k;\eta) = \frac{g^2}{12\pi^2}\Big[\ln^2\big( {-k\eta}  \big) - 2 \ln\big( {-k\eta} \big)\,\ln\big( {-k\eta_0}\big)  \Big] \ee where $\mathcal{P}_0(k)$ is the power spectrum in absence of environmental coupling, $g$ is the effective coupling to the environmental fields and $\eta_0$ is a renormalization scale chosen as the beginning of the de Sitter stage when modes of cosmological relevance today were deeply sub-Hubble. The power spectrum is suppressed at large scales with a concomitant violation of scale invariance. This result confirms those obtained with the quantum master equation in ref.\cite{boydensmat} in an independent and complementary framework. The coupling of the inflaton field to sub-Hubble degrees of freedom may be responsible for dissipative effects and \emph{may} provide an explanation for the large scale suppression of the power spectrum in the CMB. The case of nearly massless environmental scalar fields minimally coupled to gravity yield stronger corrections enhanced by infrared effects, although these environmental fields may be severely constrained by CMB observations.

In the case of environmental fields with masses $\gg H$ we obtain a local ``Fermi'' limit of the effective action to leading order in $H/M$ leading to a very weakly self-coupled inflaton, and show that dissipative effects arise at next to leading order.

\section{The model:}\label{model}
We consider a spatially flat Friedmann-Robertson-Walker (FRW)
cosmological space-time and two interacting scalar fields $\phi,\varphi$ with a generic interaction of the form $\lambda \, \phi^m\,\varphi^n$.  The field $\phi$ is an inflaton-like scalar field minimally coupled to gravity, this is the ``system'', and the field $\varphi$  is the ``environment'',  this field is traced out (integrated over) in the time evolution of the density matrix  and its correlation functions determine the \emph{influence action} contribution in the non-equilibrium effective action. We study the following relevant cases  : i) $m=1,n=2$ and $\varphi$ is  massless and conformally coupled to gravity, this case models degrees of freedom that remain sub-Hubble throughout inflation, ii) $m=1,n=2$ and $\varphi$ is light with $M_\varphi/H \ll 1$ and minimally coupled, this case exhibits strong secular and infrared enhancement, but may be constrained by their contribution to isorcurvature or entropy perturbations, iii) $m=2,n=1$ and $\varphi$ is  massive with $M_\varphi \gg H$ and minimally coupled, this case yields a local ``Fermi'' limit of the effective action with dissipative corrections that are suppressed by powers of $H/M_{\varphi}$.

 In comoving
coordinates, the action is given by
\bea
S & = &\int d^3x \; dt \;  a^3(t) \Bigg\{
\frac{1}{2}{\dot{\phi}^2}-\frac{(\nabla
\phi)^2}{2a^2}-\frac{1}{2}\Big(M^2_\phi+\xi_\phi \; \mathcal{R}\Big)\phi^2  \nonumber \\ & + &
\frac{1}{2}{\dot{\varphi}^2}-\frac{(\nabla
\varphi)^2}{2a^2}-\frac{1}{2}\Big(M^2_\varphi+\xi_\varphi \; \mathcal{R}\Big)\varphi^2   -  \lambda \,  \phi^m\,:\varphi^n:  \Bigg\}\,, \label{lagrads}
\eea
with \be \mathcal{R} = 6 \left(
\frac{\ddot{a}}{a}+\frac{\dot{a}^2}{a^2}\right) \ee being the Ricci
scalar,   $\xi=
0,1/6$ correspond  to minimal and
conformal coupling respectively. We consider de Sitter space time with $a(t) = e^{H t}$.

 The interaction has been normal-ordered
\be :\varphi^n: = \varphi^n- \langle \varphi^n \rangle \label{no} \ee where the brackets $\langle (\cdots)\rangle$ refer to the expectation value in the initial density matrix (see below).

We will be primarily focused on two special cases for the interactions:
\begin{itemize}
\item  \textbf{   m=1, n=2 :} For this case we will consider the field $\phi$ to be massive with $M_\phi \ll H$ and minimally coupled to gravity,   the influence action from tracing over the environmental field $\varphi$ will be studied in two relevant cases: i) $\varphi$ is massless and conformally coupled to gravity. These environmental fields effectively describe degrees of freedom that remain sub-Hubble throughout inflation. This case was studied in ref.\cite{boydensmat} in terms of the time evolution of the  quantum density matrix via the master equation and we will compare the results of the influence action approach to those obtained in this reference.   This is the simplest case that yields a stochastic Langevin description with    Gaussian but colored noise and a one-loop self-energy correction to the equations of motion of the inflaton fluctuation.
     ii) $\varphi$ is light with $M_\varphi \ll H$ and minimally coupled to gravity. This case exhibits stronger infrared and secular divergences, but their relevance may be severely constrained by  observations that do not support  isocurvature or entropy perturbations.   The correlation function of the stochastic noise and the self-energy obey a de Sitter space time analog of a fluctuation-dissipation relation (see discussion in section (\ref{sec:stocha})).

    \item  \textbf{  m=2, n=1 :} in this case we will consider the environmental field $\varphi$ to be in the principal series of the de Sitter group $SO(3,1)$ with $M^2_\varphi \gg H^2$\cite{marolf}. This case yields a local effective field theory for the super Hubble fluctuations of the system field $\phi$ in a consistent expansion in $H/M_\varphi \ll 1$ akin to a local ``Fermi'' theory in Minkowski space time (see discussion in section (\ref{sec:fermi})) and dissipative corrections are suppressed by powers of $H/M_{\varphi}$.

\end{itemize}

It is convenient to pass to conformal time $\eta = -e^{-Ht}/H$ with
\be  a(t(\eta)) = -\frac{1}{H\eta} \,,\label{aofeta}\ee
  and introduce a conformal rescaling of the fields
\begin{equation}
 \phi(\vx,t) = \frac{\chi(\vx,\eta)}{a(t(\eta))}~~;~~ \varphi(\vx,t) = \frac{\psi(\vx,\eta)}{a(t(\eta))}.\label{rescale}
\end{equation}
After discarding surface terms the action becomes   \be  S    =
  \int d^3x \; d\eta \, \Big\{\mathcal{L}_0[\chi]+\mathcal{L}_0[\psi]+\mathcal{L}_I[\chi,\psi] \Big\} \; \label{rescalagds}\ee
  where
  \bea \mathcal{L}_0[\chi] & = & \frac12\left[
{\chi'}^2-(\nabla \chi)^2-\mathcal{M}^2_{\chi}(\eta) \; \chi^2 \right] \label{l0chi}\\
\mathcal{L}_0[\psi] & = & \frac12\left[{\psi'}^2-(\nabla \psi)^2-\mathcal{M}^2_{\psi}(\eta) \; \psi^2  \right] \label{l0psi}\\ \mathcal{L}_I[\chi,\psi] & = & -\frac{\lambda}{(-H\eta)^p} \;  \chi^m\,:\psi^n:    ~~;~~ p= 4-m-n \; , \label{lI}\eea

with primes denoting derivatives with respect to
conformal time $\eta$ and
\be
\mathcal{M}^2_{\chi}(\eta)  = \Big[\frac{M^2_{\phi}}{H^2}+12\Big(\xi_{\phi} -
\frac{1}{6}\Big)\Big]\frac{1}{\eta^2}  ~~ ,  ~~ \mathcal{M}^2_{\psi}(\eta)  = \Big[\frac{M^2_{\varphi}}{H^2}+12\Big(\xi_{\varphi} -
\frac{1}{6}\Big)\Big]\frac{1}{\eta^2}  , \label{massds2}
\ee   respectively. Since the coupling $\lambda$ has dimensions of (mass)$^p$ we will consider the weak coupling case with $\lambda/H^p \ll 1$. To simplify notation later we write the interaction Lagrangian density as
\be \mathcal{L}_I[\chi,\psi] = -g \,J[\chi]\,\mathcal{O}[\psi] \label{LIsimple} \ee with
\bea g & = & \frac{\lambda}{H^p} \label{dimlessg} \\
J[\chi] & \equiv &  {(\chi(\vx,\eta))^m}  \label{Jchi} \\
\mathcal{O}[\psi] & \equiv & \frac{:(\psi(\vx,\eta))^n:}{{(-\eta)^p}} \label{Opsi}
\eea the coupling $g$ is dimensionless and the normal ordering prescription (\ref{no}) leads to
\be \langle \mathcal{O}[\psi] \rangle = 0 \,, \label{zerovev}\ee where the average  is in the initial density matrix (see below).

In the non-interacting case $g =0$ the Heisenberg equations of motion for the spatial Fourier modes of wavevector $\vec{k}$ for the conformally rescaled fields are
\bea
&& \chi''_{\vk}(\eta)+
\Big[k^2-\frac{1}{\eta^2}\Big(\nu^2_\chi -\frac{1}{4} \Big)
\Big]\chi_{\vk}(\eta)  =   0    \label{chimodes}\\
&& \psi''_{\vk}(\eta)+
\Big[k^2-\frac{1}{\eta^2}\Big(\nu^2_\psi -\frac{1}{4} \Big)
\Big]\psi_{\vk}(\eta)  =   0  \label{psimodes}\eea
where
\be \nu^2_{\chi} = \frac{9}{4}- \Big(\frac{M^2_{\phi}}{H^2}+12\, \xi_\phi \Big) ~~;~~  \nu^2_{\psi} = \frac{9}{4}- \Big(\frac{M^2_{\varphi}}{H^2}+12\, \xi_\varphi \Big) \,.
\label{nusa} \ee

In this and next sections we will focus on the fields $\phi,\varphi$   with $\nu^2 >0$, namely
\be \frac{9}{4} > \Big(\frac{M^2 }{H^2}+12\, \xi  \Big) \ee with  $\xi = 0,1/6$ for minimal or conformal coupling respectively and obtain the generic form of the influence action.

The Heisenberg fields are expanded in a comoving volume $V$ as
\bea
\chi(\vx,\eta) & = & \frac{1}{\sqrt{V}}\,\sum_{\vq} \Big[b_{\vq}\,g(q,\eta)+ b^\dagger_{-\vq}\,g^*(q,\eta) \Big]\,e^{i\vq\cdot\vx} \label{chiex} \\
\psi(\vx,\eta) & = & \frac{1}{\sqrt{V}}\,\sum_{\vk} \Big[a_{\vq}\,u(q,\eta)+ a^\dagger_{-\vq}\,u^*(q,\eta) \Big]\,e^{i\vk\cdot\vx}\,. \label{psiex} \eea
We choose Bunch-Davies conditions for both fields, namely
\be b_{\vq} |0\rangle_{\chi} =0 ~~;~~ a_{\vk} |0\rangle_{\psi} =0 \label{bdvac}\ee
and
\bea
 && g(q,\eta)= \frac{1}{2}\,e^{i\frac{\pi}{2}(\nu_\chi+\frac{1}{2})}\,\sqrt{-\pi\,\eta}\,H^{(1)}_{\nu_\chi}(-q\eta)\label{gqeta}\\&&
 u(k,\eta)= \frac{1}{2}\,e^{i\frac{\pi}{2}(\nu_\psi+\frac{1}{2})}\,\sqrt{-\pi\,\eta}\,H^{(1)}_{\nu_\psi}(-k\eta)\,.\label{uketa}\eea Quantization with non-Bunch Davies boundary conditions can be studied similarly with straightforward generalizations in terms of arbitrary linear combinations of the mode functions (\ref{gqeta},\ref{uketa}),  here we consider this simpler case to highlight the main physical consequences. The $\chi$ field  is considered to be  minimally coupled, $\xi_\phi =0$ and nearly massless with $M_\phi/H \ll1$, from which it follows that as $-q\eta \rightarrow 0$
 \be g(q,\eta) \propto 1/\eta \,.\label{smaleta}\ee This behavior in the super-Hubble limit will lead to strong secular contributions in the long time limit and is the hallmark of the classicalization of the quantum fluctuations, which are dominated by a growing mode\cite{polarski}. We will  obtain  the influence action up to second order in the system-environment coupling $\lambda$. In the case when the interaction is of the form $\chi :\psi^2:$  the influence action to lowest order is determined by a one-loop correlation function of the environmental fields. This case yields a stochastic description in terms of a Langevin equation with a one-loop self-energy and a Gaussian noise whose correlation function is determined by the one-loop correlator of the environmental fields.

 We will also study the interaction $\chi^2\,\psi$ with the environmental field in this case being minimally coupled ($\xi_\varphi=0$) and  with $M^2/H^2 \gg 9/4  $ corresponding to $\nu = i\mu$ with real $\mu$. This case yields a local ``Fermi'' limit for the effective action. Minimally coupled scalar fields with mass $\gg H$  belong to the principal series representation of the de Sitter group $SO(3,1)$\cite{marolf}  and require  a slight modification of the mode function treatment. This case will be studied in detail in section (\ref{sec:fermi}).

 The time evolution of a   density matrix initially prepared at time $\eta_0$ is given by
 \be \rho(\eta)= U(\eta,\eta_0)\,\rho(\eta_0)\,U^{-1}(\eta,\eta_0) \,,\label{rhoeta}\ee where $\mathrm{Tr}[\rho(\eta_0)]=1$ and  $U(\eta,\eta_0)$ is the unitary time evolution of the full theory, it obeys
\be i\frac{d}{d\eta} U(\eta,\eta_0) = H(\eta) \,U(\eta,\eta_0)~~;~~ U(\eta_0,\eta_0) =1 \label{U} \ee where $H(\eta)$ is the total Hamiltonian. Therefore
\be U(\eta,\eta_0) = \mathbf{T}\Big[e^{-i\int_{\eta_0}^\eta H(\eta')d\eta'}\Big]~~;~~  U^{-1}(\eta,\eta_0) = \widetilde{\mathbf{T}}\Big[e^{i\int_{\eta_0}^\eta H(\eta')d\eta'}\Big] \label{Uofeta}\ee with $\mathbf{T}$ the time-ordering symbol describing evolution forward in time and $\widetilde{\mathbf{T}}$ the anti-time ordered symbol describing evolution backwards in time.

Consider the initial density matrix at a conformal time $\eta_0$ and for the conformally rescaled fields
  to be of the form
\begin{equation}
 {\rho}(\eta_0) =  {\rho}_{\chi}(\eta_0) \otimes
 {\rho}_{\psi}(\eta_0) \,.\label{inidensmtx}
\end{equation} This choice while ubiquitous in the literature neglects possible initial correlations, we will adopt this choice with the understanding that the role of initial correlations between the system and environment remains to be studied more deeply.

The initial time $\eta_0$ is chosen so that all the modes of the inflaton-like field that are of cosmological relevance today are deeply sub-Hubble at the initial time. Since we are considering a (\emph{nearly}) de Sitter space-time, this initial time must be earlier than or equal to the time at which the slow-roll stage begins (we discuss this point in section (\ref{sec:stocha}) below).

Our goal is to evolve this initial density matrix in (conformal) time obtaining (\ref{rhoeta}) and trace over the environmental degrees of freedom $\varphi$ ($\psi$) leading to a \emph{reduced} density matrix for the system degrees of freedom $\chi$ namely
\be \rho^r_\chi(\eta) = \mathrm{Tr}_{\psi} \rho(\eta)\,. \label{rhofired}\ee

  Furthermore, there is no natural choice of the initial density matrices for the system or environmental fields, therefore to exhibit the main physical consequences of tracing over the environmental degrees of freedom in the simplest setting we choose
\be  {\rho}_{\chi}(\eta_0) = |0\rangle_{\chi}\,{}_\chi\langle 0|~~;~~  {\rho}_{\psi}(\eta_0) = |0\rangle_{\psi}\,{}_\psi\langle 0| \,,\label{BDinirho} \ee namely both fields are in their respective Bunch-Davies vacua. This condition can be generalized straightforwardly.  With this choice  the normal ordering prescription in (\ref{rescalagds}) becomes
 \be :(\psi(\vx,\eta))^n: = (\psi(\vx,\eta))^n - {}_\psi\langle 0|(\psi(\vx,\eta))^n|0\rangle_\psi \,. \label{norord}\ee

 In the field basis the matrix elements of $ {\rho}_{\chi}(\eta_0);{\rho}_{\psi}(\eta_0)$
are given by
\begin{equation}
\langle \chi | {\rho}_{\chi}(\eta_0) | \chi'\rangle =
\rho_{\chi,0}(\chi ,\chi')~~;~~\langle \psi | {\rho}_{\psi}(\eta_0) | \psi'\rangle =
\rho_{\psi,0}(\psi ;\psi')\,, \label{fieldbasis}
\end{equation} and we have suppressed the coordinate arguments of the fields in the matrix elements. In this basis
\bea   \rho(\chi_f,\psi_f;\chi'_f,\psi'_f;\eta) & = &      \langle \chi_f;\psi_f|U(\eta,\eta_0) {\rho}(0)U^{-1}(\eta,\eta_0)|\chi'_f;\psi'_f\rangle \nonumber \\
& = & \int D\chi_i D\psi_i D\chi'_i D\psi'_i ~ \langle \chi_f;\psi_f|U(\eta,\eta_0)|\chi_i;\psi_i\rangle\,\rho_{\chi,0}(\chi_i;\chi'_i)\times \nonumber\\
&& \rho_{\psi,0}(\psi_i;\psi'_i)\,
 \langle \chi'_i;\psi'_i|U^{-1}(\eta,\eta_0)|\chi'_f;\psi'_f\rangle \label{evolrhot}\eea The $\int D\chi$ etc, are functional integrals where the spatial argument has been suppressed. The matrix elements of the   forward and backward time evolution operators can be written as path integrals, namely
 \bea   \langle \chi_f;\psi_f|U(\eta,\eta_0)|\chi_i;\psi_i\rangle  & = &    \int \mathcal{D}\chi^+ \mathcal{D}\psi^+\, e^{i \int^\eta_{\eta_0} d\eta' d^3 x \mathcal{L}[\chi^+,\psi^+]}\label{piforward}\\
 \langle \chi'_i;\psi'_i|U^{-1}(\eta,\eta_0)|\chi'_f;\psi'_f\rangle &  =  &   \int \mathcal{D}\chi^- \mathcal{D}\psi^-\, e^{-i \int^\eta_{\eta_0}\int d^3 x \mathcal{L}[\chi^-,\psi^-]}\label{piback}
 \eea where
 $ \mathcal{L}[\chi,\psi] $ can be read off (\ref{rescalagds})   and
 the boundary conditions on the path integrals are
  \bea     \chi^+(\vec{x},\eta_0)=\chi_i(\vec{x})~;~
 \chi^+(\vec{x},\eta)  &  =  &   \chi_f(\vec{x})\,,\nonumber \\   \psi^+(\vec{x},\eta_0)=\psi_i(\vec{x})~;~
 \psi^+(\vec{x},\eta) & = & \psi_f(\vec{x}) \,,\label{piforwardbc}\\
     \chi^-(\vec{x},\eta_0)=\chi'_i(\vec{x})~;~
 \chi^-(\vec{x},\eta) &  = &    \chi'_f(\vec{x})\,,\nonumber \\   \psi^-(\vec{x},\eta_0)=\psi'_i(\vec{x})~;~
 \psi^-(\vec{x},\eta) & = & \psi'_f(\vec{x}) \,.\label{pibackbc} \\
 \eea

 The fields $\chi^\pm,\psi^\pm$ describe the time evolution forward   ($+$) with $U(\eta,\eta_0)$  and backward  ($-$ ) with $U^{-1}(\eta,\eta_0)$  as befits the time evolution of a density matrix. This is the Schwinger-Keldysh formulation\cite{schwinger,keldysh,maha} of time evolution of density matrices.

 As usual one can linearly couple sources $h^\pm$  to the fields $\chi^\pm$ on the forward ($+$) and backward ($-$)  branches, so that functional derivatives with respect to these sources yield the correlation functions along the time branches and mixed correlation functions for example
 \bea \langle \chi^+(\eta) \chi^+(\eta') \rangle & = &  \mathrm{Tr}\,\mathbf{T}(\chi(\eta) \chi(\eta')) \,\rho \label{plusplus}\\ \langle \chi^-(\eta) \chi^-(\eta') \rangle & = &  \mathrm{Tr}\,\widetilde{\mathbf{T}}(\chi (\eta) \chi (\eta')) \,\rho \label{minusminus}  \\
 \langle \chi^+(\eta)\chi^-(\eta') \rangle & = & \mathrm{Tr}\,(\chi(\eta')\chi(\eta)\,\rho \label{plusminus}\\  \langle \chi^-(\eta)\chi^+(\eta') \rangle & = & \mathrm{Tr}\,(\chi(\eta)\chi(\eta')\,\rho \,.\label{minusplus}\eea

 We are particularly interested in the power spectrum of inflaton fluctuations given by the \emph{equal time correlation function}
 \be P(k,\eta) =\frac{ k^3}{2\pi^2}\,\langle \phi_{\vk}(\eta)\phi_{-\vk}(\eta) \rangle = \frac{ k^3 H^2 \eta^2 }{2\pi^2}\,\langle \chi_{\vk}(\eta)\chi_{-\vk}(\eta) \rangle\,, \label{powspec}\ee the equal time  average can be written in terms of a ``center of mass'' combination
 \be \widetilde{\chi}_{\vk} = \frac{1}{2} \big(\chi^+_{\vk}+ \chi^-_{\vk}\big) \,,\label{cm1}\ee it is straightforward to confirm that
 \be \langle \chi_{\vk}(\eta)\chi_{-\vk}(\eta) \rangle = \langle \widetilde{\chi}_{\vk}(\eta) \widetilde{\chi}_{-\vk}(\eta) \rangle \,.  \label{cmid}\ee This a consequence of the fact that at equal times, the time and anti-time ordered correlation functions coincide with the Wightmann functions (\ref{plusminus},\ref{minusplus}). This result will be useful below to obtain the power spectrum from the effective action.

 The reduced density matrix for the light field $\chi$ is obtained by tracing over the bath ($\psi$) variables, namely
\be \rho^{r}(\chi_f,\chi'_f;\eta) = \int D\psi_f \,\rho(\chi_f,\psi_f;\chi'_f,\psi_f;\eta) \,,\label{rhored} \ee we find
\be \rho^{r}(\chi_f,\chi'_f;\eta) = \int D\chi_i   D\chi'_i  \,  \mathcal{T}[\chi_f,\chi'_f;\chi_i,\chi'_i;\eta;\eta_0] \,\rho_\chi(\chi_i,\chi'_i;\eta_0)\,,\label{rhochieta} \ee
where the time evolution kernel is given by the following path integral representation
\be \mathcal{T}[\chi_f,\chi'_f;\chi_i,\chi'_i;\eta;\eta_0] = {\int} \mathcal{D}\chi^+ \,  \mathcal{D}\chi^- \, e^{i S_{eff}[\chi^+,\chi^-;\eta]} \ee  where  the total effective action that yields the time evolution of the reduced density matrix is
 \be S_{eff}[\chi^+,\chi^-;\eta] = \int^\eta_{\eta_0} d\eta' \int d^3x \Big[\mathcal{L}_0[\chi^+]- \mathcal{L}_0[\chi^-] \Big] + \mathcal{F}[J^+, J^-]\,, \label{Seff}\ee
with
the following boundary conditions on the forward ($\chi^+$) and backward  ($\chi^-$) path integrals
\bea &  &   \chi^+(\vec{x},\eta_0)=\chi_i(\vec{x})~;~
 \chi^+(\vec{x},\eta)  =   \chi_f(\vec{x}) \nonumber \\
&  &   \chi^-(\vec{x},\eta_0)=\chi'_i(\vec{x})~;~
 \chi^-(\vec{x},\eta)  =   \chi'_f(\vec{x}) \,.\label{bcfipm}\eea   $\mathcal{F}[J^+;J^-]$    is the \emph{influence action} where $J^\pm \equiv J[\chi^\pm]$  are defined in eqn. (\ref{Jchi}),  it is determined by the trace over the environmental fields   and  is given by
 \be  e^{i\mathcal{F}[J^+;J^-]}   =   \int D\psi_i  \, D\psi'_i D\psi_f  \,\rho_{\psi}(\psi_i,\psi'_i;\eta_0) \, \int \mathcal{D}\psi^+ \mathcal{D}\psi^- \, e^{i   \int d^4x \Big\{\left[\mathcal{L}_+[\psi^+;J^+]-\mathcal{L}_-[\psi^-;J^-] \right] \Big\}}\label{inffunc}\ee where we used the shorthand notation
 \be \mathcal{L}_{\pm}[\psi^\pm;J^\pm] = \mathcal{L}_0[\psi^\pm]-g J[\chi^\pm]\mathcal{O}[\psi^\pm] ~~;~~  x \equiv (\eta, \vx) ~~;~~ \int d^4 x  \equiv \int^\eta_{\eta_0} d\eta' \int d^3x \,,\label{d4x} \ee and
 the boundary conditions on the path integrals are
 \be \psi^+(\vec{x},\eta_0)=\psi_i(\vec{x})~;~
 \psi^+(\vec{x},\eta)=\psi_f(\vec{x})~~;~~ \psi^-(\vec{x},\eta_0)=\psi'_i(\vec{x})~;~
 \psi^-(\vec{x},\eta)=\psi_f(\vec{x}) \,. \label{bcchis} \ee

In (\ref{inffunc}), $J[\chi^\pm]$ act  as   \emph{external sources} coupled to the composite operator $\mathcal{O}(\psi)$, therefore, it is clear that
\be e^{i\mathcal{F}[J^+;J^-]} = \mathrm{Tr}_{ \psi } \Big[ \mathcal{U}(\eta,\eta_0;J^+)\,\rho_\psi(\eta_0)\,  \mathcal{U}^{-1}(\eta,\eta_0;J^-) \Big]\,, \label{trasa}\ee where $J^\pm\equiv J[\chi^\pm]$ and $\mathcal{U}(\eta,\eta_0;J^\pm)$ is the   time evolution operator in the $\psi$ sector in presence of \emph{external sources} $J^\pm$ namely \be \mathcal{U}(\eta,\eta_0;J^+) = \mathbf{T}\Big( e^{-i \int_{\eta_0}^\eta H_\psi[J^+(\eta')]d\eta'}\Big) ~~;~~
\mathcal{U}^{-1}(\eta,\eta_0;J^-) = \widetilde{\mathbf{T}}\Big( e^{i \int_{\eta_0}^\eta H_\psi[J^-(\eta')]d\eta'}\Big) \ee
where
\be H_\psi[J^\pm(\eta)] = H_{0 \psi}(\eta)+g\,\int d^3x J[\chi^\pm(\eta)]\mathcal{O}(\psi) \label{timevchi}\ee and $\widetilde{\mathbf{T}}$ is the \emph{anti-time ordered evolution operator} describing  evolution backwards in time. In (\ref{timevchi}) $H_{0\psi}(\eta)$ is the free field Hamiltonian for the field $\psi$ which depends explicity on time as a consequence of the $\eta$ dependence of $\mathcal{M}^2_{\psi}(\eta)$ in (\ref{massds2}) and in the interaction term $J[\chi^\pm]$ are \emph{classical} c-number sources.

 The calculation of the influence action is facilitated by passing to the interaction picture for the Hamiltonian $H_\psi[J(\eta)]$, defining
\be  \mathcal{U}(\eta;\eta_0;J^\pm) = \mathcal{U}_0(\eta;\eta_0) ~ \mathcal{U}_{ip}(\eta;\eta_0;J^\pm) \label{ipicture} \ee where $\mathcal{U}_0(\eta;\eta_0)$ is the time evolution operator of the free field $\psi$ (namely for $g=0$) and cancels out in the trace in (\ref{trasa}), and
\be \mathcal{U}_{ip}(\eta;\eta_0;J^\pm) \simeq 1 -i g \int^\eta_{\eta_0} d\eta'\int d^3 x J[\chi^\pm(\vx,\eta')]\mathcal{O}(\psi(\vx,\eta'))+ \cdots \label{Uip}\ee and $\psi(\vx,\eta)$ has the time evolution of free fields given by (\ref{psimodes}).

  Now the trace can be obtained systematically in perturbation theory in $g$. Up to $\mathcal{O}(g^2)$  in the cumulant expansion  we find (using the shorthand notation (\ref{d4x})) \bea \mathcal{F}[J^+,J^-] & = &    - g \int d^4x \Big( J^+[x]-J^-[x]\Big)\,\langle \mathcal{O}(x)\rangle_{\psi} \nonumber + \\ & & \frac{i g^2 }{2} \int d^4x_1 \int d^4x_2 \Bigg\{ J^+[x_1]\,J^+[x_2]\,G_c^{++}(x_1;x_2)+ J^-[x_1]\,J^-[x_2]\,G_c^{--}(x_1;x_2) \nonumber \\
 & - & J^+[x_1]\,J^-[x_2]\,G_c^{+-}(x_1;x_2)- J^-[x_1]\,J^+[x_2]\,G_c^{-+}(x_1;x_2)\Bigg\}\,. \label{finF}\eea In this expression $J^\pm[x]\equiv J[\chi^\pm(\vx,\eta)]$, and the \emph{connected} correlation functions are given by
\begin{eqnarray}
&& G_c^{-+}(x_1;x_2) =   \langle
{\cal O}(x_1) {\cal O}(x_2)\rangle_{ \psi } - \langle \mathcal{O}(x_1)\rangle_{ \psi} \langle \mathcal{O}(x_2)\rangle_{ \psi} =   {G}_c^>(x_1;x_2) \,,\label{ggreat} \\&&  G_c^{+-}(x_1;x_2) =   \langle
{\cal O}(x_2) {\cal O}(x_1)\rangle_{ \psi} - \langle \mathcal{O}(x_2)\rangle_{ \psi}  \langle \mathcal{O}(x_1)\rangle_{\{\psi\}} =   {G}_c^<(x_1;x_2)\,,\label{lesser} \\&& G_c^{++}(x_1;x_2)
  =
{ G}_c^>(x_1;x_2)\Theta(\eta_1-\eta_2)+ {  G}_c^<(x_1;x_2)\Theta(\eta_2-\eta_1) \,,\label{timeordered} \\&& G_c^{--}(x_1;x_2)
  =
{ G}_c^>(x_1;x_2)\Theta(\eta_2-\eta_1)+ {  G}_c^<(x_1;x_2)\Theta(\eta_1-\eta_2)\,,\label{antitimeordered}
\end{eqnarray} in terms of interaction picture fields, where
\be \langle (\cdots) \rangle_{ \psi} = \mathrm{Tr}_{ \psi}(\cdots)\rho_\psi(\eta_0)\,. \label{expec}\ee
Furthermore, for the case of hermitian operators $\mathcal{O}$ as considered here it follows that
\be G_c^>(x_1;x_2) = G_c^<(x_2;x_1)\,, \label{ident}\ee and the normal ordering prescription (\ref{Opsi}) leads to $\langle \mathcal{O}(x)\rangle =0$ (see \ref{zerovev}). Writing (\ref{finF}) in terms of $G^>,G^<$, using the property (\ref{ident}) and following the steps detailed in ref.(\cite{boyeff}) we find
\bea \mathcal{F}[J^+, J^-]  &  = &  i\,g^2\int d^3x_1 d^3x_2 \int^\eta_{\eta_0} d\eta_1\,\int^{\eta_1}_{\eta_0} d\eta_2\,\Bigg\{ J^+(\vx_1,\eta_1)J^+(\vx_2,\eta_2)\,G^>(x_1;x_2)   \nonumber \\ & + &  J^-(\vx_1,\eta_1)J^-(\vx_2,\eta_2)\,G^<(x_1;x_2)  -   J^+(\vx_1,\eta_1)J^-(\vx_2,\eta_2)\,G^<(x_1;x_2)  \nonumber\\
  &- &   J^-(\vx_1,\eta_1)J^+(\vx_2,\eta_2)\,G^>(x_1;x_2)\Bigg\} ~~;~~ x_1 = (\eta_1,\vx_1) ~~ \mathrm{etc}\,.\label{Funravel}\eea

  In a spatially flat FRW cosmology, spatial translational invariance implies that
  \be G^{\lessgtr}(x_1,x_2) = G^{\lessgtr}(\vx_1-\vx_2;\eta_1,\eta_2) \equiv \frac{1}{V} \sum_{\vp} \mathcal{K}^\lessgtr_{p}(\eta_1,\eta_2)\,e^{-i\vp\cdot(\vx_1-\vx_2)} \,,\label{kernelsft}\ee and in terms of the spatial Fourier transforms
  \be J[\chi^\pm(\vx,\eta)] \equiv \frac{1}{\sqrt{V}}\sum_{\vk} J^\pm_{\vk}(\eta) \,e^{i\vk\cdot\vx}\,,\label{fts}\ee we find the general form of the influence action up to second order in the coupling,
 \bea \mathcal{F}[J^+, J^-]  &  = &  i\,g^2\,\sum_{\vk} \int^\eta_{\eta_0} d\eta_1\,\int^{\eta_1}_{\eta_0} d\eta_2\,\Bigg\{ J^+_{\vk}(\eta_1)J^+_{-\vk}(\eta_2)\,\mathcal{K}^>_{k}(\eta_1;\eta_2)     +   J^-_{\vk}(\eta_1)J^-_{-\vk}(\eta_2)\,\mathcal{K}^<_{k}(\eta_1;\eta_2) \nonumber \\ & - &    J^+_{\vk}(\eta_1)J^-_{-\vk}(\eta_2)\,\mathcal{K}_{k}^<(\eta_1;\eta_2)  -  J^-_{\vk}(\eta_1)J^+_{-\vk}(\eta_2)\,\mathcal{K}^>_{k}(\eta_1;\eta_2)\Bigg\}  \,.\label{Funfina}\eea

 \vspace{2mm}

 \section{Stochastic effective action}\label{sec:stocha}
\begin{figure}[ht!]
\begin{center}
\includegraphics[height=3.5in,width=3.5in,keepaspectratio=true]{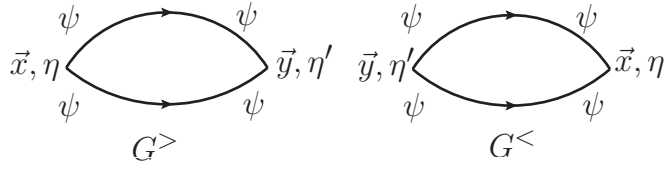}
\caption{The correlation functions $G^>(\vx-\vy,\eta,\eta'),G^<(\vx-\vy,\eta,\eta') $. }
\label{fig:oneloop}
\end{center}
\end{figure}

 In this section we consider the   case of $m=1,n=2$ namely
 \be J[\chi] =  {\chi(\vx,\eta)}  ~~;~~ \mathcal{O}[\psi] = \frac{:(\psi(\vx,\eta))^2:}{{(-\eta)}} \,,\label{simplecase} \ee
 For this case we find
 \bea && G^>(\vx-\vy,\eta,\eta') = \frac{1}{\eta\eta'}~ \mathrm{Tr}_{ \psi }[:\psi^2(\vx,\eta):\,:\psi^2({\vy},\eta'):\,\rho_{\psi}(\eta_0)] \,,\label{ggreatcor} \\
&& G^<(\vx-\vy,\eta,\eta') = \frac{1}{\eta\eta'}~ \mathrm{Tr}_{ \psi }[:\psi^2(\vy,\eta'):\,:\psi^2(\vx,\eta):\,\rho_{\psi}(\eta_0)]\,, \label{gless} \eea
 and
 \bea && \mathcal{K}^>[q;\eta,\eta'] \equiv K[q;\eta,\eta'] =\frac{ 2}{\eta\eta'} \int \frac{d^3k}{(2\pi)^3}\,u(k,\eta)u^*(k,\eta')u(p,\eta)u^*(p,\eta')~~;~~ p=|\vk+\vq|\nonumber \\
&& \mathcal{K}^<(q;\eta,\eta')= \mathcal{K}^>[q;\eta',\eta]= K^*[q;\eta,\eta'] \,,\label{kernels} \eea where $u(k,\eta)$, are the mode functions (\ref{uketa}). These correlation functions describe a one-loop contribution from integrating out the environmental degrees of freedom displayed in fig.(\ref{fig:oneloop}).

The interpretation of the effective action (\ref{Seff}) becomes clear by introducing the center of mass and relative variables
\be \widetilde{\chi}(\vx,\eta) = \frac{1}{2} \big(\chi^+(\vx,\eta)+\chi^-(\vx,\eta)\big) ~~;~~ R(\vx,\eta)= \big(\chi^+(\vx,\eta)-\chi^-(\vx,\eta)\big)\,, \label{cmrelvas}\ee in terms of which and passing to spatial Fourier transforms we find
\begin{eqnarray}
iS_{eff}[\widetilde{\chi},R] & = & \int_{\eta_0}^{\eta} d\eta' \sum_{\vec k} \left\{-i R_{-\vec k}(\eta')\left(\widetilde{\chi}^{''}_{\vec k}(\eta')+\Ok^2(\eta')\,\widetilde{\chi}_{\vk}(\eta') \right)\right\} \nonumber \\
                 & - & \int_{\eta_0}^{\eta} d\eta_1 \int_{\eta_0}^{\eta} d\eta_2 \left\{\frac{1}{2}\,R_{\vec k}(\eta_1)\,{\mathcal{N} }_k(\eta_1;\eta_2)\,R_{-\vec k}(\eta_2) + R_{-\vec k}(\eta_1)\,
i\Sigma^R_k(\eta_1;\eta_2)\, \widetilde{\chi}_{\vec k}(\eta) \right \} \nonumber \\
& + & \int d^3x  R_i(\vx){\widetilde{\chi}}'(\vx,\eta_0) \,,
\label{effac}
\end{eqnarray} where
\be \Ok^2(\eta) =  \Big[k^2-\frac{1}{\eta^2}\Big(\nu^2_\chi -\frac{1}{4} \Big)
\Big]\,. \label{omegak}\ee  In the last term (surface term) in (\ref{effac}) we have neglected a contribution from $R(\vx,\eta)$ since this vanishes upon taking the trace to obtain expectation values or correlation functions. The kernels $\mathcal{N},\Sigma$ in (\ref{effac}) are given by
\be {\mathcal{N} }_k(\eta_1;\eta_2) = \frac{{g^2}}{2} \Big[\mathcal{K}^>_{k}(\eta_1;\eta_2)+ \mathcal{K}^<_{k}(\eta_1;\eta_2)\big] =  {g^2}  \,\mathrm{Re}[K(k;\eta_1;\eta_2)]\,, \label{Nker}\ee
\bea  \Sigma^R_k(\eta_1;\eta_2) & = &   {-ig^2}  \Big[\mathcal{K}^>_{k}(\eta_1;\eta_2)- \mathcal{K}^<_{k}(\eta_1;\eta_2)\big]\Theta(\eta_1-\eta_2) \nonumber \\ & = &    {2 g^2} \,\mathrm{Im}[K(k;\eta_1;\eta_2)]\Theta(\eta_1-\eta_2)\,. \label{sigmaker}\eea

These  are the de Sitter space time analog of the generalized fluctuation-dissipation relation in Minkowski space-time\cite{boyeff}.

The term quadratic in $R$ in (\ref{effac}) can be written in terms of a Gaussian noise variable, namely
\be
  \exp\Big\{-\frac{1}{2} \int d\eta_1 \int d\eta_2 R_{-\vec k}(\eta_1)\mathcal{N}_k(\eta_1;\eta_2)R_{\vec k}(\eta_2)\Big\} = \int {\cal D}\xi \,\mathcal{P}\big[\xi\big] \, e^{i \int d\eta' \, \xi_{-\vec k}(\eta') R_{\vec k}(\eta')  } \label{quadR} \ee where
  \be \mathcal{P}\big[\xi\big] =
\exp\Big\{-\frac{1}{2} \int d\eta_1 \int d\eta_2 ~~ \xi_{\vec k}(\eta_1)
\mathcal{N}^{-1}_k(\eta_1;\eta_2)\xi_{-\vec k}(\eta_2) \Big\} \,. \label{probaxi}
\ee

In terms of the center of mass and relative variables the initial density matrix  in (\ref{rhochieta}) becomes
\be \rho_\chi(\chi_i,\chi'_i;\eta_0) = \rho_\chi(\widetilde{\chi}_i+\frac{R_i}{2},\widetilde{\chi}_i-\frac{R_i}{2};\eta_0) \label{rhochicmrel}\ee therefore it proves convenient to perform a  Wigner transform, namely

\begin{equation}
{\cal W}(\widetilde{\chi}_i ; \Pi_i) = \int D R_i e^{-i\int d^3x \Pi_i(\vec x)
R_i(\vec x)} \rho_{\chi,0}(\widetilde{\chi}_i+\frac{R_i}{2};\widetilde{\chi}_i-\frac{R_i}{2})\,, \label{wignerrho}\ee whose inverse transform yields
\be \rho_{\chi,0}(\widetilde{\chi}_i+\frac{R_i}{2};\widetilde{\chi}_i-\frac{R_i}{2})= \int D \Pi_i~
e^{i\int d^3x \Pi_i(\vec x) R_i(\vec x)} {\cal W}(\widetilde{\chi}_i;\Pi_i)\,.\label{inverwignerrho} \end{equation} Gathering all the above steps leads to
\be \rho^{r}(\chi_f,\chi'_f;\eta) =  \int D\widetilde{\chi}_i D\Pi_i \int \mathcal{D}\widetilde{\chi} \mathcal{D}R \mathcal{D}\xi ~\mathcal{P}[\xi]~e^{iS_{eff}[\widetilde{\chi},R,\xi;\eta]}~   {\cal W}(\widetilde{\chi}_i;\Pi_i) \,, \label{rhorfini}\ee where
\be  S_{eff}[\widetilde{\chi},R,\xi;\eta]  =  -\int_{\eta_0}^{\eta} d\eta_1 \sum_{\vk} R_{-\vec k}(\eta_1) \left[{\widetilde{\chi}}^{''}_{\vec k}(\eta_1)+\Ok^2(\eta_1){\widetilde{\chi}}_{\vec k}(\eta_1)+\int_{\eta_0}^{\eta_1} d\eta_2 ~~ \Sigma^{R}_k(\eta_1;\eta_2){\widetilde{\chi}}_{\vec k}(\eta_2)-\xi_{\vec k}(\eta_1) \right]\,. \label{Seffstocha}\ee  and we carried out the functional integral over $R_i$, which is effectively a delta function yielding $ {\widetilde{\chi}}'(\eta_0) = \Pi_i $.  Obviously the effective action describes a stochastic process, the path integral over the relative variable $R$ in (\ref{rhorfini}) yields a functional delta function
\be \delta\Bigg[{\widetilde{\chi}}^{''}_{\vec k}(\eta)+\Ok^2(\eta){\widetilde{\chi}}_{\vec k}(\eta)+\int_{\eta_0}^{\eta} d\eta_1 ~~ \Sigma^{R}_k(\eta;\eta'){\widetilde{\chi}}_{\vec k}(\eta')-\xi_{\vec k}(\eta) \Bigg] \label{deom} \ee whose solution is the \emph{Langevin equation}

\be  {\widetilde{\chi}}^{''}_{\vec k}(\eta)+\Ok^2(\eta){\widetilde{\chi}}_{\vec k}(\eta)+\int_{\eta_0}^{\eta} d\eta_1 ~~ \Sigma^{R}_k(\eta;\eta_1){\widetilde{\chi}}_{\vec k}(\eta_1)=\xi_{\vec k}(\eta)\,.  \label{langevin}\ee The noise $\xi_{\vk}(\eta)$ is Gaussian and colored with the correlation function
\be  \overline{ \xi_{\vk}(\eta_1)\xi_{-\vk'}(\eta_2)}   \equiv   \frac{\int \mathcal{D}\xi~\mathcal{P}[\xi] ~~\xi_{\vk}(\eta_1) \xi_{-\vk'}(\eta_2) }{\int \mathcal{D}\xi~\mathcal{P}[\xi]  } = \mathcal{N}_k(\eta_1;\eta_2)\, \delta_{\vk,\vk'} ~~;~~   \overline{\xi_{\vk}(\eta) }   =0 \,,\label{noiseav}\ee where $\mathcal{N}$ is given by eqn. (\ref{Nker}). As mentioned above (see the discussion after (\ref{Nker},\ref{sigmaker})) the noise correlation function $\mathcal{N}$ and the self energy $\Sigma$ obey a de Sitter space time generalization of the fluctuation dissipation relation. This formulation is akin to the path integral framework for \emph{classical stochastic} field theories developed by Martin-Rose and Siggia\cite{msr}.

It is clear from (\ref{noiseav}) that formally $\xi \propto g$  since $\overline{\xi \xi} = \mathcal{N} \propto g^2$, an observation that will be relevant in the analysis below.

The solution of the Langevin equation (\ref{langevin}) is a function(al) of the noise term and the initial conditions $\chi_{\vk}(\eta_0);\chi'_{\vk}(\eta_0)=\Pi_{\vk,i}$,  to obtain correlation functions and observable quantities  associated with the $\chi$ fields two different averages are involved\cite{boyeff}:
 \begin{itemize}
 \item{ Average over the initial conditions $\chi_{\vk}(\eta_0);\chi'_{\vk}(\eta_0)$ with the initial density matrix $\rho_{\chi}(\eta_0)$,   we refer to these averages simply as
     \be \langle (\cdots ) \rangle_{\chi} = \mathrm{Tr}_{\chi}(\cdots) \rho_{\chi}(\eta_0) \,. \label{traceavera}\ee}

\item {Average over the noise, this is   a Gaussian average with the probability distribution function $\mathcal{P}[\xi]$ with first and second moments given by eqn. (\ref{noiseav}), these averages are referred to as
    \be \overline{(\cdots)}   \equiv   \frac{\int \mathcal{D}\xi~\mathcal{P}[\xi] ~~ (\cdots) }{\int \mathcal{D}\xi~\mathcal{P}[\xi]  } \,. \label{avegausnois}\ee}

   \item{Therefore the \emph{total} average of correlation functions is given by
   \be \overline{\langle C[\chi;\xi;\eta]\rangle_\chi }=   \frac{\int \mathcal{D}\xi~\mathcal{P}[\xi] ~~ \mathrm{Tr}_\chi(C[\chi;\xi;\eta]\rho_{\chi}(\eta_0)) }{\int \mathcal{D}\xi~\mathcal{P}[\xi]  } \,. \label{fullaverage}\ee }

\end{itemize}

In principle one should obtain the Wigner distribution function as in eqn. (\ref{wignerrho}) however this is just a representation of the initial density matrix $\rho_{\chi,0}$,   in the field basis passing to a Wigner transform in the relative coordinate. Therefore the averages over the initial conditions are averages with the initial density matrix, which in this article is taken to be a pure density matrix describing the (free) $\chi$ fields in their Bunch-Davies vacuum.

 While this program may be unfamiliar a simple example illustrates how to implement it. Let us first consider the non-interacting case    $g=0$ in which case both the noise and self-energy terms vanish (these are related by the generalization of the fluctuation-dissipation relation). The Langevin equation (\ref{langevin}) now simply becomes the field equations (\ref{chimodes}) whose solutions are linear combinations of the mode function $g(k,\eta),g^*(k,\eta)$ given by eqn. (\ref{gqeta}). Since our main goal is to extract the corrections to the power spectrum and the growing mode in the super Hubble limit, it is more convenient to expand the solution in terms of the real growing ($g_+$)  and decaying modes ($g_-$), namely\cite{boydensmat}
\be \widetilde{\chi}_{\vk}(\eta) = Q_{\vk} \, g_+(k,\eta) + P_{\vk}\, g_-(k;\eta) \label{chiQP} \ee with
\be   g_+(k;\eta) = \sqrt{\frac{-\pi\eta}{2}}~ Y_{\nu_\chi}(-k\eta) ~~;~~ g_-(k;\eta) = \sqrt{\frac{-\pi\eta}{2}}~ J_{\nu_\chi}(-k\eta)\,, \label{gpms}\ee where $Y,J$ are Bessel functions. In the super Hubble limit $-k\eta \rightarrow 0$
 \be Y_{\nu_\chi}(-k\eta) \propto (-k\eta)^{-\nu_\chi}~~;~~ J_{\nu_\chi}(-k\eta) \propto (-k\eta)^{ \nu_\chi}  \,. \label{suphub}\ee
   In terms of the operators $b_{\vk},b^\dagger_{\vk}$ in the expansion (\ref{noiseav}) it follows that
   \bea Q_{\vk} & =  & \frac{1}{\sqrt{2}} \,\Big( b_{\vk}\,e^{i\frac{\pi}{2}(\nu_\chi+\frac{3}{2})}+ b^\dagger_{-\vk}\,e^{-i\frac{\pi}{2}(\nu_\chi+\frac{3}{2})} \Big) ~~;~~ Q^\dagger_{\vk} = Q_{-\vk} \label{Qs} \\  P_{\vk}
&  = &  \frac{i}{\sqrt{2}} \,\Big( b^\dagger_{-\vk}\,e^{-i\frac{\pi}{2}(\nu_\chi+\frac{3}{2})}  - b_{\vk}\,e^{i\frac{\pi}{2}(\nu_\chi+\frac{3}{2})}\Big)~~;~~ P^\dagger_{\vk} = P_{-\vk}\,. \label{Ps} \eea As discussed in ref.\cite{boydensmat},   $Q_{\vk},P_{\vk}$ are canonical conjugate variables. In the Heisenberg picture the operators $Q_{\vk},P_{\vk}$ feature the following expectation values in the initial density matrix
 \bea &&  \langle Q_{\vk} \rangle = Tr_{\chi}  Q_{\vk} {\rho}_{\chi}(\eta_0)= 0 ~~;~~ \langle P_{\vk} \rangle = Tr_{\chi} P_{\vk}{\rho}_{\chi}(\eta_0)=0 \nonumber\\ &&  \langle Q_{\vk}  Q_{-\vk'}\rangle = Tr_{\chi} Q_{\vk} Q_{-\vk'} {\rho}_{\chi}(\eta_0)= \frac{1}{2} \,\delta_{\vk,\vk'} ~~;~~  \langle P_{\vk}P_{-\vk'}\rangle =  Tr_\chi P_{\vk}P_{-\vk'}{\rho}_{\chi}(\eta_0)  = \frac{1}{2} \,\delta_{\vk,\vk'}\nonumber \\ && \langle Q_{\vk} P_{-\vk'} \rangle =   Tr_\chi Q_{\vk} P_{-\vk'} {\rho}_{\chi}(\eta_0) = \frac{i}{2}\, \delta_{\vk,\vk'} \,. \label{expvalini} \eea Therefore any arbitrary correlation function of the $\chi$ field can be obtained from the expansion (\ref{chiQP}), the average over initial conditions (\ref{traceavera}) are given by the (Gaussian) expectation values (\ref{expvalini}).
  In particular, from the result (\ref{powspec}) and the identity (\ref{cmid})  the unperturbed ($g=0$)   power spectrum for the massless case $M_\varphi =0$ ($\nu_\chi = 3/2$)   in the super Hubble limit $-k\eta \rightarrow 0$  follows from (\ref{powspec}, \ref{expvalini})
 \be \mathcal{P}_0(k,\eta) = \Big( \frac{H}{2\pi}\Big)^2 \,, \label{powspecff}\ee this is the usual (nearly) scale invariant result.

 Having addressed the free field case, let us now consider solving the Langevin equation (\ref{langevin}) in a perturbative expansion
 \be \widetilde{\chi}_{\vk}(\eta) = \widetilde{\chi}^{(0)}_{\vk}(\eta) + g \chi^{(1)}_{\vk}(\eta;\xi)+ \cdots \label{pt}\ee where $\widetilde{\chi}^{(0)}_{\vk}(\eta)$ is given by (\ref{chiQP}). The full solution is of the form
 \be \widetilde{\chi}_{\vk}(\eta) \equiv \chi_{\vk}[Q_{\vk},P_{\vk};\xi;\eta]\,.\label{fulsol}\ee Because the Langevin equation (\ref{langevin}) is linear, the full solution is a linear function of $Q_{\vk}, P_{\vk};\xi$ and after solving it one can obtain correlation functions by averaging over the noise with (\ref{noiseav}) and over the $Q,P's$ with (\ref{expvalini}).

\vspace{2mm}

\textbf{Modelling sub-Hubble degrees of freedom:}

\vspace{2mm}

It remains to obtain the self energy $\Sigma$ and noise correlation function $\mathcal{N}$ (\ref{sigmaker},\ref{Nker}) which are determined by the kernel $K[k;\eta;\eta']$ (\ref{kernels}). Our goal is to integrate out sub-Hubble environmental degrees of freedom to obtain the influence action for the super-Hubble fluctuations of the inflaton field. For this purpose, we now consider the environmental  field $\varphi$ ($\psi$) to be massless and conformally coupled to gravity, namely $M^2_\varphi =0~;~\xi_\varphi = 1/6$, for which $\nu_\psi=1/2$ and the mode functions in (\ref{uketa}) become
 \be u(k,\eta) = \frac{e^{-ik\eta}}{\sqrt{2k}}\,. \label{conf}\ee This is precisely the sub-Hubble limit ($-k\eta \gg 1$) of the mode functions for a massive and minimally coupled scalar field, therefore this choice for   the environmental field $\psi$ effectively models  degrees of freedom that remain sub-Hubble during inflation. We will comment later on the modifications that arise if this environmental field is chosen to be nearly massless and minimally coupled.

  With the mode functions (\ref{conf})  we find (see also ref.\cite{boydensmat})
 \be  K[k;\eta,\eta'] = -\frac{i}{8\pi^2\,\eta\eta'} \, \frac{e^{-ik(\eta-\eta')}}{(\eta-\eta'-i\varepsilon)}~~;~~ \varepsilon \rightarrow 0^+ \,,\label{kernel} \ee Therefore the kernel features a non-local and a local part,
 \be  K[k,\eta,\eta'] =   -\frac{i}{8\pi^2\eta\eta'}  {e^{-ik(\eta-\eta')}}\,\mathcal{P}\Big[\frac{1}{\eta-\eta'} \Big] + \, \frac{1}{8\pi\,\eta^2}\,\delta(\eta-\eta')\,, \label{LNL}\ee where $\mathcal{P}$ is the principal part.  It is convenient to write the principal part as
 \be \mathcal{P}\Big[\frac{1}{\eta-\eta'} \Big] = \frac{\eta-\eta'}{(\eta-\eta')^2+\varepsilon^2} =
-\frac{1}{2}\, \frac{d}{d\eta'}\,\ln\Bigg[\frac{(\eta-\eta')^2+\varepsilon^2 }{(-\eta_0)^2}\Bigg]\,. \label{PP}\ee In  (\ref{PP}) we have introduced $-\eta_0$    to render the argument of the logarithm dimensionless, this scale acts as a subtraction or renormalization scale just as in the usual renormalization program in Minkowski space-time and is chosen so as to cancel surface terms in integration by parts in the long time limit. The particular choice of this scale, coinciding with the initial time at which the density matrix is ``prepared'',  is guided by the condition that the wavelengths of cosmological relevance are deeply sub-Hubble at this time . Furthermore, as discussed below this scale also determines the scale at which the mass of the inflaton field is renormalized.  The self-energy and noise kernels are given by
\bea \Sigma^R_k(\eta;\eta')  & = &  \frac{g^2 \cos[k(\eta-\eta')]}{8\pi^2\eta\eta'} ~~ \frac{d}{d\eta'}\,\ln\Bigg[\frac{(\eta-\eta')^2+\varepsilon^2 }{(-\eta_0)^2}\Bigg] \label{sigmc}\\
\mathcal{N}_k(\eta,\eta') & = &  \frac{g^2}{8\pi \eta\eta'}\,\Big[\delta(\eta-\eta')-\frac{\sin[k(\eta-\eta')]}{\pi(\eta-\eta')}\Big]\,. \label{Nmc}\eea

\vspace{2mm}

\subsection{Solving the Langevin equation: the dynamical renormalization group.}
We begin by analyzing the solution of the   Langevin equation (\ref{langevin}) for $\xi=0$. This analysis will pave the way to understanding the full solution with the noise term. Since our main goal is to obtain the corrections to the power spectrum in the super-Hubble limit, we will focus on the long time $-\eta \rightarrow 0$  and super-Hubble $-k\eta \rightarrow 0$ limits.

As a first step let us consider the non-local contribution from the self-energy term in the Langevin equation (\ref{langevin}), with the self-energy given by (\ref{sigmc}) this term is given by
\be \int_{\eta_0}^{\eta} d\eta_1   \Sigma^{R}_k(\eta;\eta_1){\widetilde{\chi}}_{\vec k}(\eta_1) =  \frac{g^2}{8\pi^2\eta}\int^\eta_{\eta_0} d\eta' \cos[k(\eta-\eta')]\frac{\widetilde{\chi}_{\vk}(\eta')}{\eta'} \frac{d}{d\eta'}\,\ln\Bigg[\frac{(\eta-\eta')^2+\varepsilon^2 }{(-\eta_0)^2}\Bigg]\,, \ee   integrating by parts the derivative term,   the contribution from the lower limit yields
\be   - \frac{g^2}{8\pi^2\eta^2_0 } \cos[k(\eta-\eta_0)] \widetilde{\chi}_{\vk}(\eta_0) \, \ln\Bigg[\frac{(\eta-\eta_0)^2  }{(-\eta_0)^2}\Bigg]\,, \ee this contribution vanishes in the long time limit $\eta \rightarrow 0$, this is the advantage of taking $-\eta_0$ in the argument of the logarithm in (\ref{PP}). The contribution from the upper limit in the integration by parts is
\be \frac{g^2}{8\pi^2\eta^2 } \widetilde{\chi}_{\vk}(\eta)\, \ln\Big[\frac{\varepsilon^2}{(-\eta_0)^2}\Big] \,, \label{MR} \ee now the \emph{homogeneous} equation (\ref{langevin}) becomes
\bea  &&  {\widetilde{\chi}}^{''}_{\vec k}(\eta)+\Bigg[k^2+ \frac{1}{\eta^2}\Big(\frac{M^2_\phi}{H^2}+  \frac{g^2}{8\pi^2}\ln\Big[\frac{\varepsilon^2}{(-\eta_0)^2}\Big]\Big)-2 \Bigg]{\widetilde{\chi}}_{\vec k}(\eta)  \nonumber \\ & = &
 \frac{g^2}{4\pi^2\eta}\int^\eta_{\eta_0} d\eta' \, \ln\Big[\frac{(\eta-\eta') }{(-\eta_0)}\Big]\,\frac{d}{d\eta'} \Big[\cos[k(\eta-\eta')]\frac{\widetilde{\chi}_{\vk}(\eta')}{\eta'}  \Big] \,. \label{homolan} \eea In the first line of this expression we have combined $\Ok^2(\eta)$ given by (\ref{omegak})  with (\ref{MR}) to exhibit the fact that the term (\ref{MR}) is a \emph{mass renormalization}, a result that was also found in the quantum master equation approach\cite{boydensmat}. Therefore we absorb the divergent contribution (\ref{MR}) into a mass renormalization writing
 \be \frac{M^2_\phi}{H^2}+  \frac{g^2}{8\pi^2}\ln\Big[\frac{\varepsilon^2}{(-\eta_0)^2}\Big] = \frac{M^2_{\phi,R}}{H^2}\,. \label{massren}\ee We note that $-\eta_0$ enters as a renormalization scale for the mass. This is important, identifying $\eta_0$ with a time scale of the order of or earlier than the onset of slow roll inflation the renormalized mass of the inflaton during slow roll inflation is $M^2_{\phi,R}$ and a nearly scale invariant spectrum of super-Hubble fluctuations entails $M^2_{\phi,R}/H^2 \ll 1$.

 From now on we refer to the mass as the renormalized value and
 \be \nu^2_\chi = \frac{9}{4}- \frac{M^2_{\phi,R}}{H^2}\,, \label{nuchiren}\ee and the bracket in the left hand side (\ref{homolan}) becomes $\Ok^2(\eta)$ in terms of the renormalized mass. We now proceed to solve (\ref{homolan}) in perturbation theory, anticipating the main results, such a perturbative expansion features secular terms that require a resummation program.   We will implement the \emph{dynamical renormalization group}\cite{drg,nigel} to provide a non-perturbative resummation of these secular terms.

 We begin by writing\footnote{In the homogeneous case the first order correction is of $\mathcal{O}(g^2)$.}
 \be \widetilde{\chi}_{\vk}(\eta) = \widetilde{\chi}^{(0)}_{\vk}(\eta)+ g^2 \widetilde{\chi}^{(1)}_{\vk}(\eta) + \cdots \label{ptexp}\ee leading to the hierarchy of equations
 \bea     \frac{d^2}{d\eta^2} {\widetilde{\chi}}^{(0)}_{\vec k}(\eta)+  \Ok^2(\eta)  {\widetilde{\chi}}^{(0)}_{\vec k}(\eta)  & = & 0 \label{zeroord}\\
 \frac{d^2}{d\eta^2} {\widetilde{\chi}}^{(1)}_{\vec k}(\eta)+  \Ok^2(\eta)  {\widetilde{\chi}}^{(1)}_{\vec k}(\eta) & =  & \frac{1}{4\pi^2\eta}\int^\eta_{\eta_0} d\eta' \, \ln\Big[\frac{(\eta-\eta') }{(-\eta_0)}\Big]\,\frac{d}{d\eta'} \Big[\cos[k(\eta-\eta')]\frac{\widetilde{\chi}^{(0)}_{\vk}(\eta')}{\eta'}  \Big] \nonumber \\   \vdots & = & \vdots \,. \nonumber   \eea The zeroth order solution is given by (\ref{chiQP}), namely
 \be  {\widetilde{\chi}}^{(0)}_{\vec k}(\eta) = Q_{\vk} \, g_+(k,\eta) + P_{\vk}\, g_-(k;\eta)\,,\label{0thsol} \ee inserting this solution into the first order equation it can be solved using the Green's function of the differential operator. For the full solution (\ref{0thsol}) the integral in the first order equation with the convolution with the Green's function for general values of $\nu_{\chi}$ and $k$ cannot be done analytically. However we are primarily interested in obtaining the corrections to the power spectrum   at long times and in the super-Hubble limit ($-k\eta \ll 1$)  and for the case when the field $\phi$ is nearly massless, namely $M_{\phi,R}/H \ll 1$ as this case yields a (nearly) scale invariant power spectrum in the free theory.

 These limits justify the following approximations
 \begin{itemize}
 \item in the super-Hubble limit for $M_{\phi,R}/H\rightarrow 0$ ($\nu_\chi = 3/2$) \be g_+(k;\eta) = 1/(k^{3/2}\eta)~~;~~g_-(k;\eta) = (k^{3/2}\eta^2)/3 \label{supHub}\ee

     \item neglect the decaying term in the zeroth-order solution, namely \be {\widetilde{\chi}}^{(0)}_{\vec k}(\eta) \simeq  Q_{\vk} \, g_+(k,\eta)= \frac{Q_{\vk}}{k^{3/2}\eta} \label{supHubchi} \ee

         \item the lower limit of the integral in the first order equation in (\ref{zeroord}) will be replaced by $\eta_* \simeq -1/k$, because for $\eta > \eta_*$ the integrand is dominated by the growing mode. In this interval $\cos[k(\eta-\eta')] \simeq 1$ for super Hubble wave vectors. The contribution to the integral from the interval $\eta > \eta' > \eta_*$ will be seen to feature secular growth with time, because the growing mode dominates, whereas the contribution from $\eta_* > \eta' \geq \eta_0$ remains perturbatively small, since in this region   ${\widetilde{\chi}}^{(0)}_{\vec k}(\eta') \propto 1/\sqrt{k}$ and oscillates rapidly.

 \end{itemize}

 With these approximations and to leading order in long time and  super-Hubble limits we find  that the inhomogeneity of the first order equation, namely the right hand side  in (\ref{zeroord}), is given by
 \be I(\eta) = \frac{Q_{\vk}}{4\pi^2k^{3/2}\eta^3}\, \Big[\ln\Big(\frac{\eta}{\eta_0}\Big)-1\Big] \, .\label{supHubinh}\ee  This result clearly displays the secular divergent growth as $\eta \rightarrow 0^-$. The solution of the first order  equation becomes
 \be \widetilde{\chi}^{(1)}_{\vk}(\eta) = \int^\eta_{\eta_*} G[\eta,\eta']\,I(\eta') d\eta' \label{1stordi}\ee where in the super-Hubble limit the retarded Green's function is given by
 \be  G[\eta,\eta'] \theta(\eta-\eta') = \frac{1}{3}\, \Big[\frac{\eta^2}{\eta'}- \frac{{\eta'}^{\,2}}{\eta }\Big]\,\theta(\eta-\eta') \, .\label{greensfun}\ee   The lower limit $\eta_* = -1/k$ in (\ref{1stordi}) reflects the region of integration in which the corresponding wavevector is super-Hubble and we can   use of the approximations above. The contribution from the interval $\eta_0 < \eta' < \eta_*$ is truly perturbative and does not feature secular growth. To leading order in this limit we find
 \be \widetilde{\chi}^{(1)}_{\vk}(\eta) = \Bigg(\frac{Q_{\vk}}{k^{3/2}\eta}\Bigg)\,F[\eta] \label{firsto}\ee where to leading order for $\eta/\eta_* \simeq -k\eta \rightarrow 0$ we find
 \be F[\eta] = -\frac{1}{12\pi^2} \Big\{ \frac{1}{2}\,\ln^2\Big(\frac{\eta}{\eta_*} \Big) + \ln\Big(\frac{\eta}{\eta_*}\Big)\,\ln\Big(\frac{\eta_*}{\eta_0}\Big)    \Big\}\,. \label{Feta}\ee Therefore the solution of the homogeneous equation (\ref{homolan}) in the long time limit and for super-Hubble wavelengths, keeping only the growing mode  is
 \be \widetilde{\chi}_{\vk}(\eta) = \frac{Q_{\vk}}{k^{3/2}\eta}\,\Big[ 1+ g^2F[\eta]+ \cdots \Big] \,. \label{solugrou}\ee   Obviously $F[\eta]$ features secular growth as $\eta/\eta_* \simeq -k\eta \rightarrow 0$ and the perturbative solution eventually breaks down in the asymptotic long time limit. Furthermore, the form of the solution  (\ref{solugrou}) suggests that the corrections are a renormalization of the amplitude $Q_{\vk}$. In order to obtain a solution that is asymptotically well behaved we implement the \emph{dynamical renormalization group} (DRG) resummation program\cite{drg,nigel}. We introduce a ``wave function'' renormalization $Z[\tau]$ and an arbitrary renormalization scale $\tau$ and write the amplitude $Q_{\vk}$ as
 \be Q_{\vk} = Q_{\vk}[\tau]Z[\tau] ~~;~~Z[\tau] = 1+g^2z_1[\tau] + \cdots \,.\label{wfren} \ee Inserting this expansion in the solution (\ref{solugrou}),
 \be  \widetilde{\chi}_{\vk}(\eta) = \frac{Q_{\vk}[\tau]}{k^{3/2}\eta}\,\Big[ 1+ g^2\,\big(F[\eta]+z_1[\tau]\big)+ \cdots \Big] \,. \label{soluren}\ee We now choose $z_1[\tau]$ to precisely cancel $F[\eta=\tau]$ thereby improving the perturbative expansion, with this choice the improved solution is
 \be  \widetilde{\chi}_{\vk}(\eta) = \frac{Q_{\vk}[\tau]}{k^{3/2}\eta}\,\Big[ 1+ g^2\,\big(F[\eta]-F[\tau]\big)+ \cdots \Big] \,, \label{soluren2}\ee the convergence is improved by choosing $\tau$ arbitrarily close to a fixed time $\eta$. However the solution \emph{does not} depend on the arbitrary renormalization scale $\tau$, therefore
 \be \frac{\partial \widetilde{\chi}_{\vk}(\eta)}{\partial \tau} = 0 \,, \label{RG}\ee  leading to the dynamical renormalization group equation\cite{drg,nigel}
\be \frac{d}{d\tau} Q_{\vk}[\tau] \Big[1+\cdots]- Q_{\vk}[\tau] g^2  \frac{d}{d\tau} F[\tau] = 0 \,. \label{drgeq} \ee To leading order the solution is given by
\be Q_{\vk}[\tau] = Q_{\vk}[\tau_*] \,e^{g^2\big[F[\tau]-F[\tau_*] \big]} \,.\label{drgsol}\ee  Since the scales $\tau,\tau_*$ are arbitrary, we now choose $\tau= \eta, \tau_*=\eta_*$ and since $F[\eta_*]=0$ we finally find the (DRG) improved growing mode solution of the \emph{homogeneous}  Langevin equation  in the long-time and super-Hubble limits
\be \widetilde{\chi}_{\vk}(\eta) = \frac{Q_{\vk}[\eta_*]}{k^{3/2}\eta}~ e^{\,g^2\,F[\eta]} \,. \label{drgfinasol}\ee

\vspace{1mm}

\textbf{Solution of the inhomogeneous: an alternative method.}

\vspace{1mm}

 The solution of the full inhomogeneous Langevin equation (\ref{langevin}), is the solution of the homogeneous (\ref{homolan}) plus the solution of the inhomogeous for which we would need the Green's function of the integro-differential operator on the left hand side of (\ref{langevin}). Finding the Green's function of this operator is a daunting task, below we present a method that allows to obtain an approximation to the full solution by exploiting the multiplicative renormalization arising from the (DRG) solution of the homogeneous equation studied above.

To begin with we follow the same procedure as above and integrate by parts the self-energy term absorbing the contribution of the upper limit of integration into the mass renormalization and neglect the contribution from the lower limit which vanishes in the long time limit. Secondly,  we write the noise term (inhomogeneity) in (\ref{langevin}) as
\be \xi_{\vk}(\eta) \equiv g \,\widetilde{\xi}_{\vk}(\eta) \label{noiseg}\ee to exhibit explicitly that formally the
noise is of $\mathcal{O}(g)$, (so that $\widetilde{\xi}\simeq \mathcal{O}(1)$). The Langevin equation (\ref{langevin}) becomes
\be   {\widetilde{\chi}}^{''}_{\vec k}(\eta)+\Ok^2(\eta)\,{\widetilde{\chi}}_{\vec k}(\eta)    =
 \frac{g^2}{4\pi^2\eta}\int^\eta_{\eta_0} d\eta' \, \ln\Big[\frac{(\eta-\eta') }{(-\eta_0)}\Big]\,\frac{d}{d\eta'} \Big[\cos[k(\eta-\eta')]\frac{\widetilde{\chi}_{\vk}(\eta')}{\eta'}  \Big] + g \,\widetilde{\xi}_{\vk}(\eta)\,. \label{noisilan} \ee where again $\Ok^2(\eta)$ is in terms of the renormalized mass. In the next step we exploit the multiplicative renormalization result from the (DRG) and write
 \be {\widetilde{\chi}}_{\vec k}(\eta) = \widetilde{\Psi}_{\vk}(\eta)\,e^{\alpha(\eta)} \label{multi}\ee where
 \be \alpha(\eta) = g^2 \alpha_1(\eta)+ g^3 \alpha_2(\eta) + \cdots \label{alfadef}\ee and proceed to obtain $\widetilde{\Psi}$ and $\alpha$ systematically in a resummed perturbative expansion so that ${\widetilde{\chi}}_{\vec k}(\eta)$ features a uniform  asymptotic limit. In the argument of the integral on the right hand side of (\ref{noisilan}) consider
 \be  \frac{d}{d\eta'}\Big[\cos[k(\eta-\eta')]\frac{\widetilde{\chi}_{\vk}(\eta')}{\eta'}  \Big]  = \frac{d}{d\eta'}\Big[\cos[k(\eta-\eta')]\frac{\widetilde{\Psi}_{\vk}(\eta')}{\eta'}   \Big]\,e^{\alpha(\eta')}  +    \alpha'(\eta) \Big[\cos[k(\eta-\eta')]\frac{\widetilde{\Psi}_{\vk}(\eta')}{\eta'}\,e^{\alpha(\eta')}  \Big]\,. \label{inteterm}\ee  In the second term, $\alpha' \propto g^2  $ yields a contribution of $\mathcal{O}(g^4)$ to (\ref{noisilan}) and will be neglected to leading order $\mathcal{O}(g^2)$ which is the order at which the effective action has been obtained. Furthermore let us define
 \be \zeta_{\vk}(\eta') = \int^{\eta'}_{\eta_0} \ln\Big[\frac{(\eta-\eta'') }{(-\eta_0)}\Big]\,\frac{d}{d\eta''} \Big[\cos[k(\eta-\eta'')]\frac{\widetilde{\Psi}_{\vk}(\eta'')}{\eta''}  \Big]\,d\eta'' \label{zetavar}\ee so that
 \be \frac{d}{d\eta'}\,\zeta_{\vk}(\eta') =  \ln\Big[\frac{(\eta-\eta') }{(-\eta_0)}\Big]\,\frac{d}{d\eta'} \Big[\cos[k(\eta-\eta')]\frac{\widetilde{\Psi}_{\vk}(\eta')}{\eta'}  \Big]~~;~~\zeta_{\vk}(\eta_0)=0 \,,\label{derzet}\ee therefore, keeping only the first term in (\ref{inteterm}) consistently up to $\mathcal{O}(g^2)$ the integral   on the right hand side
 of (\ref{noisilan}) is written as
 \be \int^\eta_{\eta_0} ~ e^{\alpha(\eta)}~ \frac{d}{d\eta'}\,\zeta_{\vk}(\eta') ~  d\eta' =
 e^{\alpha(\eta)} ~\zeta_{\vk}(\eta) + \mathcal{O}(g^2) \label{intbypar}\ee where we integrated by parts  and used again that $\alpha' \propto g^2$. The term $\mathcal{O}(g^2)$ yields a contribution of $\mathcal{O}(g^4)$ to the Langevin equation (\ref{noisilan}) and will be neglected to  leading order, thus the integral term becomes
 \be e^{\alpha(\eta)}~  \int^{\eta}_{\eta_0} \ln\Big[\frac{(\eta-\eta') }{(-\eta_0)}\Big]\,\frac{d}{d\eta'} \Big[\cos[k(\eta-\eta')]\frac{\widetilde{\Psi}_{\vk}(\eta')}{\eta'}  \Big]\,d\eta'  \,. \label{intzeta}\ee Since we focus on the long-time and super-Hubble limits , we anticipate that the integral is dominated by the growing mode in the interval  $\eta_* < \eta' <\eta$, we keep only this mode and    cutoff the integral at a lower limit $\eta_*$   approximating $\cos[k(\eta-\eta')] \simeq 1$.

Introducing the definition (\ref{multi}) into the left hand side   of (\ref{noisilan}) writing it in terms of $\widetilde{\Psi}$ and $\alpha$ the Langevin equation (\ref{noisilan})  becomes
\bea && \alpha^{''}(\eta)\widetilde{\Psi}_{\vk}(\eta)+2 \, \alpha^{'}(\eta)\widetilde{\Psi}^{'}_{\vk}(\eta)- \frac{g^2}{4\pi^2\eta}\int^{\eta}_{\eta_*} \ln\Big[\frac{(\eta-\eta') }{(-\eta_0)}\Big]\,\frac{d}{d\eta'} \Big[ \frac{\widetilde{\Psi}_{\vk}(\eta')}{\eta'}  \Big]\,d\eta'  \nonumber \\ && =
-\Big[\widetilde{\Psi}^{''}_{\vk}(\eta) + \Ok^2(\eta)\widetilde{\Psi}_{\vk}(\eta)-g\widetilde{\xi}(\eta)\,e^{-\alpha(\eta)} \Big] \label{eqnalf}\eea in arriving at this expression we have neglected a term $\propto (\alpha^{'}(\eta))^2 \propto g^4$ consistently with a leading order calculation. We now impose that the second line of (\ref{eqnalf}) vanishes, namely
\be
  \widetilde{\Psi}^{''}_{\vk}(\eta) + \Ok^2(\eta)\widetilde{\Psi}_{\vk}(\eta) = g\widetilde{\xi}(\eta)\,e^{-\alpha(\eta)}     \label{vanishrhs}\ee The solution of this equation is straightforward,
  \be \widetilde{\Psi}_{\vk}(\eta) = \widetilde{\Psi}^{(0)}_{\vk}(\eta)+g\, \int^\eta_{\eta_0} G_{\vk}(\eta,\eta') \,\widetilde{\xi}(\eta')\,e^{-\alpha(\eta')} d\eta' \,, \label{solupsi} \ee where  (see eqn. (\ref{chiQP}))
  \be \widetilde{\Psi}^{(0)}_{\vk}(\eta) = Q_{\vk} \, g_+(k,\eta) + P_{\vk}\, g_-(k;\eta) \label{psihomo} \ee is the solution of the homogeneous equation in terms of the growing $ g_+(k,\eta)$ and decaying $g_-(k;\eta) $ modes and $G_{\vk}(\eta,\eta')\,\theta(\eta-\eta')$ is the retarded Green's function of the free field differential operator on the left hand side of (\ref{vanishrhs}). We now introduce this solution into the non-local self-energy contribution in the first line  of (\ref{eqnalf}), since this term is already of $\mathcal{O}(g^2)$ and the second term in the solution (\ref{solupsi})  is formally of order $g$, to leading order $g^2$  we only input $\widetilde{\Psi}^{(0)}_{\vk}(\eta)$ into   (\ref{eqnalf}), therefore the equation that determines $\alpha(\eta)$ to leading order $\mathcal{O}(g^2)$ now becomes
\be  \alpha^{''}_1(\eta)\widetilde{\Psi}^{(0)}_{\vk}(\eta)+2 \, \alpha^{'}_1(\eta)(\widetilde{\Psi}^{(0)})^{'}_{\vk}(\eta)=  \frac{1}{4\pi^2\eta}\,\int^{\eta}_{\eta_*} \ln\Big[\frac{(\eta-\eta') }{(-\eta_0)}\Big]\,\frac{d}{d\eta'} \Big[ \frac{\widetilde{\Psi}^{(0)}_{\vk}(\eta')}{\eta'}  \Big]\,d\eta' \label{finalfaen}\,. \ee Consistently with the long-time and super-Hubble limit we only keep the growing mode in $\widetilde{\Psi}^{(0)}_{\vk}\simeq  {Q_{\vk}}/{(k^{3/2}\eta)} $ and the right hand side of (\ref{finalfaen}) becomes
\[  \frac{ g^2\,Q_{\vk}}{4\pi^2k^{3/2}\eta^3}\, \Big[\ln\Big(\frac{\eta}{\eta_0}\Big)-1\Big] \,, \]
  therefore to order $g^2$ in the expansion of  $\alpha(\eta)$ (\ref{multi}) we obtain
  \be \alpha^{''}_1 - \frac{2}{\eta}\,\alpha^{'}_1  = \frac{1}{4\pi^2\,\eta^2}\Big[ \ln\big(\frac{\eta}{\eta_0} \big) -1\Big]\,, \label{eqnalfa1}\ee neglecting subdominant terms in the long time limit the solution  is given by
  \be  \alpha_1(\eta)= \alpha_1(\eta_*)+F[\eta]  \,. \label{alfa1solu}\ee where $F[\eta]$ (\ref{Feta}) had been obtained from (DRG) resummation above and $\alpha_1(\eta_*)$ is an integration  constant, it arises from the integration region $\eta_0 \leq \eta' < \eta_*$ which is non-secular and features a finite limit as $\eta \rightarrow 0$. Thus we obtain to leading order
  \be \alpha(\eta) = g^2\,\big(\alpha_1(\eta_*)+F[\eta]\big)\,. \label{alfafinag2}\ee Inserting this solution into (\ref{multi}) with (\ref{solupsi}), keeping only the growing mode, and setting the renormalized mass to zero ( $\nu_{\chi}=3/2$) we find in the long-time and super-Hubble limit
  \be \widetilde{\chi}_{\vk}(\eta) = \frac{Q_{\vk}[\eta_*]}{k^{3/2}\eta}~ e^{\,g^2\,F[\eta]}  + g\, \int^\eta_{\eta_0} G[\eta,\eta'] \,\widetilde{\xi}(\eta')\,e^{g^2\Big(F[\eta]-F[\eta']\Big)} d\eta' \,, \label{soluchifina} \ee where
  \be Q_{\vk}[\eta_*] = Q_{\vk}~ e^{g^2\alpha_1(\eta_*)} \,,\label{Qstar2}\ee and
  $G[\eta,\eta']$ is given by (\ref{greensfun}) in the super-Hubble limit. This is the DRG improved solution of the Langevin equation to leading order in the long-time and super Hubble limits, setting the noise term to zero we recover the homogeneous solution (\ref{drgfinasol}) a check of consistency with the DRG resummation.

  Now we have all the elements necessary to obtain the power spectrum. From (\ref{powspec},\ref{cmid}) and the averages over the initial phase space variables and noise (see eqns. (\ref{traceavera}-\ref{fullaverage})) it is given by
  \be P(k,\eta) =\frac{ k^3}{2\pi^2}\,\langle \phi_{\vk}(\eta)\phi_{-\vk}(\eta) \rangle = \frac{ k^3 H^2 \eta^2 }{2\pi^2}\,\,\overline{\langle \widetilde{\chi}_{\vk}(\eta)\widetilde{\chi}_{-\vk}(\eta) \rangle}\,, \label{powspecfina}\ee the super-Hubble and long time limit yield
  \bea  P(k,\eta)  & = &  \frac{H^2  }{2\pi^2}~e^{2g^2F[\eta]}~\Bigg[\langle Q_{\vk}[\eta_*]Q_{-\vk}[\eta_*]\rangle \nonumber \\ & + &    k^3\,\eta^2\, \int^\eta_{\eta_0} d\eta_1 \int^\eta_{\eta_0} d\eta_2 \, G[\eta,\eta_1]G[\eta,\eta_2] ~e^{-g^2\,\big[F[\eta_1]+F[\eta_2]\big]} \,\mathcal{N}_{k}(\eta_1;\eta_2) \Bigg]\label{Powpow}\eea where $\mathcal{N}_{k}(\eta,\eta')$ is given by (\ref{Nmc}).
  The average in the first term requires the evolution of the homogeneous equation (\ref{homolan}) for the growing mode from the initial time $\eta_0$ up to the time $\eta_* \simeq -1/k$, in turn this requires a detailed assessment of the self-energy term on the right hand side of  (\ref{homolan}) evaluated  at $\eta_*$. Although the integral is daunting even in lowest order in perturbation theory, in the interval $\eta_0 < \eta' <\eta_*$ the wavelength of the particular fluctuation is sub Hubble, $-k\eta' \gg 1$ and $\widetilde{\chi}_{\vk}(\eta') \simeq e^{-i k\eta'}/\sqrt{2k}$ which oscillates rapidly, this rapid oscillation along with that from the cosine term leads to a bound contribution which is non-secular. This is because the mode function is bounded and does not grow while the wavevector is sub Hubble. Therefore the contribution from the self-energy at the scale $\eta_*$ is perturbatively small of $\mathcal{O}(g^2)$ and non-secular leading us to conclude that
  \be \langle Q_{\vk}[\eta_*]Q_{-\vk}[\eta_*]\rangle \simeq \frac{1}{2} + \mathcal{O}(g^2) \,.\label{Qstar} \ee It remains to obtain the contribution from the noise. From the result for the noise kernel (\ref{Nmc})   we approximate $\mathcal{N}$ in the super Hubble limit by  \be \mathcal{N}_{k}(\eta_1;\eta_2) \simeq  \frac{g^2}{8\pi \eta_1\eta_2}\, \delta(\eta_1-\eta_2) \,. \label{suphubNker} \ee  Therefore the contribution from the noise term to the power spectrum (\ref{Powpow}) is given by
  \be \Big( \frac{H^2}{2\pi^2}\Big) \,\frac{g^2}{72\pi}\, k^3\,\eta^2\, \int^\eta_{\eta_*} \Bigg[\frac{\eta^2}{\eta^2_1} - \frac{\eta_1}{\eta} \Bigg]^2\,e^{2g^2\big(F[\eta]-F[\eta_1]\big)} \,d\eta_1 \,,\label{noisecontri}  \ee where we have taken the super Hubble limit and used (\ref{greensfun}) for the Green's function in this limit. The integral in (\ref{noisecontri}) cannot be done in closed form, however we can assess if it yields a substantial contribution by neglecting the exponential term which is formally $< 1$. \emph{If} the resulting integral features secular growing terms in the long-time limit, the contribution of the exponential is relevant and must be included because it damps out the secular growth at long time. We find in the long time limit
  \be  k^3\,\eta^2\, \int^\eta_{\eta_*} \Bigg[\frac{\eta^2}{\eta^2_1} - \frac{\eta_1}{\eta} \Bigg]^2\, d\eta_1 \simeq  \frac{1}{3} + 2 (-k\eta)^3\,\ln[-k\eta] \,,\label{noisecontriap}  \ee where we used $\eta_* \simeq -1/k$. Therefore in the super Hubble and long time limit this contribution approaches a \emph{perturbatively small} constant and is subleading. This result is in agreement with that obtained from the quantum master equation in ref.\cite{boydensmat} where the contribution from the local $\delta(\eta-\eta')$ term is found to be subleading.

  Therefore restoring the exponential this argument indicates that the noise contribution is perturbatively small and does not feature secular growth, the long time  and super Hubble limit of the power spectrum is completely determined by the first term in (\ref{Powpow}), namely
  \be \mathcal{P}(k,\eta) = \Big(\frac{H}{2\pi}\Big)^2 e^{-\gamma(\eta)}~(1+ \mathcal{O}(g^2))   ~~;~~\gamma(k;\eta) = \frac{g^2}{12\pi^2}\Big[\ln^2\big( {-k\eta}  \big) - 2 \ln\big( {-k\eta} \big)\,\ln\big( {-k\eta_0}\big)  \Big]\,.\label{Powpowfinal} \ee This result agrees with that obtained in ref.\cite{boydensmat} with the quantum master equation. The dependence on $\eta_0$ is a consequence of the choice of renormalization scale for the mass (see eqn.\ref{massren}), in ref.\cite{boydensmat} this aspect is discussed in detail and shown that renormalizing the mass at  a different scale $\widetilde{\eta}$ leads to replacing $\eta_0 \rightarrow \widetilde{\eta}$ in (\ref{Powpowfinal}) as a consequence of the invariance under choice of renormalization scale. The choice of $\eta_0$ as the renormalization scale and as a time scale at which slow roll inflation begins is therefore consistent with renormalizing the inflaton mass to vanish at this scale since during slow roll $M_{R,\phi} \ll H$ for the power spectrum to be scale invariant.

  Since the wavevector $k$ is considered to be sub-Hubble at the initial time $-k\eta_0 \gg 1$ and in the super Hubble limit $-k\eta \rightarrow 0^+$ it is clear that $\gamma(\eta) > 0$ and the power spectrum is \emph{suppressed} in the super Hubble limit. If $\eta_0$ is taken to be the beginning of the slow roll stage and the wavevector $k$ corresponds to a cosmologically relevant scale that became super Hubble about $8-10$  e-folds before the end of inflation at $\eta_{f}$, and if inflation lasts the minimum $60$ efolds then
  \be \ln[-k\eta_0] \simeq 50~;~\ln[-k\eta_f] \simeq -10 \Rightarrow \gamma(\eta_f) \simeq 8\,g^2 \label{dampfac}\ee then with $g = \lambda/H \sim 0.1$ the damping of the power spectrum is $\approx 10\,\%$ for modes that re-enter the Hubble radius during recombination (assuming no further corrections from the post-inflationary stage). \emph{Perhaps} this could be an explanation for the suppression of the CMB power spectrum at large scales.

  Although this result is in complete agreement with that obtained in ref.\cite{boydensmat} we have learned that the noise term is subleading and the leading contribution to the suppression of the  power spectrum arises from the ``dissipative'' self-energy kernel. In the quantum master equation approach\cite{boydensmat} it was recognized that the local contribution is subleading, therefore we identify the contribution from the noise kernel with the local contribution in the quantum master equation.

  Beyond confirming and extending the results from the quantum master equation in a complementary and independent manner, the important aspect of the results is that the interaction with   sub-Hubble fluctuations of environmental fields lead to ``dissipative'' effects and a \emph{suppression of the power spectrum at large scales}.

\vspace{2mm}

\subsection{Minimally coupled nearly massless environmental fields:} Let us now consider environmental fields that are minimally coupled to gravity, and nearly massless with $M_\varphi/H \ll 1$. Unlike the conformally coupled situation, now super Hubble environmental fluctuations are amplified and grow. Loop corrections with these fields lead to both secular and infrared enhanced contributions.  The mode functions for these fields are given by (\ref{uketa}) with
 \be \nu_\psi \simeq \frac{3}{2} - \Delta  ~~;~~ \Delta  = \frac{M^2_\varphi}{3H^2} \label{delsi}\,. \ee
 The integral in (\ref{kernels}) feature logarithmic infrared divergences for $M_\varphi/H \rightarrow 0$ which are manifest as poles in $\Delta$\cite{boyan}. Detailed analysis of these divergences shows that these arise from the regions of integration corresponding the super Hubble wave vectors and are completely determined by the growing modes.

 The calculation of the kernel $K[q,\eta,\eta']$ follows the same steps described in detail in ref. \cite{boyan} (see the appendix in the second reference). The regions of integration $k \simeq 0~;~|\vk+\vq| \simeq 0$ are isolated and an infrared cutoff $\mu$ is introduced in these regions, within which the  small argument expansion $u(p,\eta) \propto p^{-\nu_\psi}$ is used and the integration yields poles in $\Delta$. Outside these infrared regions it is safe to take $\nu_\psi = 3/2$, the details are available in the second reference in \cite{boyan}. Using the results of this reference we find to leading and next to leading  order in $\Delta$ the dissipative (self-energy) and noise kernels become
         \bea \Sigma^R_{k}(\eta,\eta')   & \equiv &    \Sigma^R_{k,IR}(\eta,\eta')+\Sigma^R_{k,cc}(\eta,\eta')\,, \label{totsig}\\ \mathcal{N}_{k}(\eta,\eta')   & \equiv &    \mathcal{N}_{k,IR}(\eta,\eta')+\mathcal{N}_{k,cc}(\eta,\eta')\label{totnoiss}\,,
         \eea where
        \be   \Sigma^R_{k,IR}(\eta,\eta')   =    \frac{g^2}{8\pi (\eta \eta')^{3/2}\,\Delta} ~\big[k^2\eta\eta']^\Delta ~ \Big[J_{\nu_\psi}(-k\eta)\,J_{\nu_\psi}(-k\eta')+Y_{\nu_\psi}(-k\eta)\,Y_{\nu_\psi}(-k\eta')\Big]\label{sigIR}\ee
        \be   \mathcal{N}_{k,IR}(\eta,\eta')   =    \frac{g^2}{8\pi (\eta \eta')^{3/2}\,\Delta} ~\big[k^2\eta\eta']^\Delta \Big[Y_{\nu_\psi}(-k\eta)\,J_{\nu_\psi}(-k\eta')-J_{\nu_\psi}(-k\eta)\,Y_{\nu_\psi}(-k\eta')\Big]\label{noisIR}\,\ee
        and $\Sigma^R_{k,cc}(\eta,\eta')~,~\mathcal{N}_{k,cc}(\eta,\eta')$ are given by eqns. (\ref{sigmc},\ref{Nmc}) respectively. We see that in this case even for super Hubble fluctuations the noise kernel is \emph{non-local} and strongly infrared and secularly enhanced (compare the powers of $\eta,\eta'$ with the local term in the conformally coupled case).

        The contributions $\Sigma^R_{k,IR}(\eta,\eta')~,~\mathcal{N}_{k,IR}(\eta,\eta')$ are a consequence of  the infrared enhancement of a minimally coupled, \emph{nearly massless} scalar field, arising from the infrared contribution of \emph{super Hubble} modes in the loop of the environmental scalar field  in the integral in (\ref{kernels}). The contributions $\Sigma^R_{k,cc}(\eta,\eta')~,~\mathcal{N}_{k,cc}(\eta,\eta')$ arise from the short wavelength modes in the loop that remain sub-Hubble and are, therefore, the same as in the conformally coupled case.

      Inspection of the different contributions available in ref.\cite{boyan} reveals the infrared enhanced terms (IR) arise  from the regions $k\simeq 0~;~ |\vk+\vq| \simeq 0$ in the integral (\ref{kernels}) whereas the latter contribution    arises precisely from the terms $e^{-ik\eta}/\sqrt{2k}$ in the mode functions $u(k;\eta)$ confirming their origin as sub-Hubble contributions. This can be understood in the case with $M_\varphi=0~;~\nu_\varphi = 3/2$, for which
      \be u(k;\eta) = \frac{e^{-ik\eta}}{\sqrt{2k}}\Big[1-\frac{i}{k\eta}
       \Big]\,, \ee when input in the loop integral in the  kernel $K[q,\eta,\eta']$ (\ref{kernels}) for example the term
       \be u(k;\eta)u^*(k;\eta')= \frac{e^{-ik(\eta-\eta')}}{2k}\Big[1-\frac{i}{k\eta}+\frac{i}{k\eta'}+\frac{1}{k^2\eta\eta'} \Big] \label{produs}\ee the first term in the bracket yields the conformally coupled result, the other terms yield infrared enhanced contributions that are cutoff by the mass and lead to the poles in $\Delta$ (see detailed discussion in ref.\cite{boyan}).  In the super Hubble limit the infrared enhanced contributions yield \be  \Sigma^R_{k,IR}(\eta,\eta')   \simeq  \frac{g^2}{4\pi^2 \,\Delta} ~\frac{\big[k^2\eta\eta']^{2\Delta}}{(k\eta\eta')^3} \,, \label{sHsigIR}\ee

      \be \mathcal{N}_{k,IR}(\eta,\eta')   \simeq   -\frac{g^2}{8\pi^2 (\eta \eta')^{3/2}\,\Delta} ~\big[k^2\eta\eta']^\Delta ~\Big[\Big(\frac{\eta'}{\eta}\Big)^{\nu_\psi}- \Big(\frac{\eta}{\eta'}\Big)^{\nu_\psi}\Big] \,. \label{sHNIR}\ee

      At this point we can implement the technical aspects of the dynamical renormalization group to the homogeneous and inhomogeneous equation, the resulting integrals are obviously more complicated   and the non-local contribution to the noise kernel and the power spectrum  is more difficult to assess. Undoubtedly the infrared and secular enhancement will lead to a stronger suppression of  the power spectrum both as a consequence of the stronger secular growth and also because of the $1/\Delta$ infrared enhancement.
      However quite aside from the technical difficulties, there are two inherently important aspects to be understood before engaging in the more technical aspects: i) minimally coupled and nearly massless environmental fields are a source of super Hubble fluctuations that are amplified, therefore they may be a source of isocurvature or entropy perturbations, which are severely constrained by the observations of the CMB, ii) our goal is to understand how degrees of freedom whose fluctuations with wavevectors that are sub Hubble all throughout inflation influence the dynamics of super Hubble fluctuations that seed the CMB upon horizon re-entry during recombination. The infrared enhanced contributions arise precisely from integration regions in the loop contributions in which  modes   become super Hubble.
      One can cutoff the integrals with a sliding physical momentum scale and integrate only the sub-Hubble components, although this would introduce an explicit time dependence through the cutoff scheme, it is clear from the analysis that the loop integration over the short wavelength modes yields a contribution that is effectively described by the conformally coupled massless fields and yields a suppression of the power spectrum. We will not pursue further the analysis of this case, simply observing that \emph{if} there are nearly massless and minimally coupled scalar fields that couple to inflationary perturbations these \emph{may} yield large suppression of the power spectrum. If future observations   allow a small component of isocurvature/entropy perturbations thus suggesting the possibility of extra scalar fields minimally coupled to gravity, then a deeper study of their impact on the power spectrum may be warranted.

      \section{Heavy fields: the ``Fermi'' limit.}\label{sec:fermi}

      We have considered the ``environmental fields'' as light in the sense that $M_{\varphi}/H \ll 1$, in this section we study the coupling of the light inflaton-like field to a \emph{heavy} scalar with $M_{\varphi}/H \gg 1$, and we focus on the interaction (see eqn. (\ref{rescalagds},\ref{lI})
\be \mathcal{L}_I[\chi,\psi]   =     {g} J[\chi]~\mathcal{O}[\psi] ~~;~~ J[\chi] = {\chi^2(\vx,\eta)}  ~~;~~\mathcal{O}(\psi) = \frac{\psi(\vx,\eta)}{(-\eta)}\label{lincoup} \ee the main reason for studying this particular coupling is because integrating out the environmental field $\psi$ in this case yields a typical $\chi^4$ type coupling in a local ``Fermi'' limit. Our goal is to study,   if and how a local Fermi limit emerges in de Sitter space time, and if and how dissipative effects are manifest in the effective action.

For the case  $M_{\varphi}/H \gg 1$ it follows from eqn. (\ref{nusa}) that $\nu_\psi $ is purely imaginary and there is a subtlety in defining the mode functions $u(k,\eta)$ in eqn. (\ref{uketa})  so that they obey Bunch-Davies boundary conditions. We define
\be \mu \equiv \frac{M_\varphi}{H} \gg 1 \ee  and following ref.(\cite{nisthandbook})   introduce the real functions
\bea \widetilde{J}_\mu(z) & = &  \mathrm{sech}\big[\frac{\pi\mu}{2}\big]~\mathrm{Re}[J_{i\mu}(z)] \label{tilJ} \\
     \widetilde{Y}_\mu(z) & = &  \mathrm{sech}\big[\frac{\pi\mu}{2}\big]~\mathrm{Re}[Y_{i\mu}(z)] \label{tilY}\eea along with
\be \widetilde{H}^{(1)}_\mu(z) = \widetilde{J}_\mu(z)+ i\widetilde{Y}_\mu(z)~~;~~\widetilde{H}^{(2)}_\mu(z) = \widetilde{J}_\mu(z)- i\widetilde{Y}_\mu(z) \,.\label{tilHs}\ee These functions feature the following asymptotic behavior for $|z|\gg \mu$
\be \widetilde{H}^{(1)}_\mu(z) \rightarrow e^{iz}\,e^{-i\pi/4}\,\sqrt{\frac{2}{\pi z}} \,. \label{asitilH}\ee The mode functions that satisfy the Bunch-Davies boundary conditions are
\be u(k,\eta) =
\frac{1}{2}\,\sqrt{-\pi \eta}\,e^{i\pi/4}\,\widetilde{H}^{(1)}_\mu(-k\eta)\,, \label{uketalarma}\ee in the super Hubble limit $-k\eta \rightarrow 0$ and for $\mu \gg 1$ we find
\be u(k,\eta)\rightarrow  \,e^{i\pi/4}\,e^{-i\gamma_\mu}\, \sqrt{\frac{- \eta}{2\mu}}~ \Big(\frac{-k\eta}{2} \Big)^{i\mu}\,,\label{suphubuk}\ee where $\gamma_\mu$ is real\cite{nisthandbook}.

We study the emergence of a local ``Fermi'' limit directly from the influence action (\ref{Funfina}). The corresponding Green's functions defined in eqn.(\ref{ggreat},\ref{lesser}) are given by
\be G^{\lessgtr}(x_1,x_2) = G^{\lessgtr}(\vx_1-\vx_2;\eta_1,\eta_2) \equiv \frac{1}{V} \sum_{\vk} \mathcal{K}^\lessgtr_{k}(\eta_1,\eta_2)\,e^{-i\vk\cdot(\vx_1-\vx_2)}  \label{gis}\ee with
\be \mathcal{K}^>_{k}(\eta_1,\eta_2) = \frac{u(k,\eta_1)\,u^*(k,\eta_2)}{\eta_1\eta_2}~~;~~\mathcal{K}^<_{k}(\eta_1,\eta_2) = \Big[\mathcal{K}^>_{k}(\eta_1,\eta_2)\Big]^* \,.\label{bigKs}\ee The influence action (\ref{Funfina}) now becomes
\bea   \mathcal{F}[J^+, J^-]     & = &    i\,g^2\,\sum_{\vk} \int^\eta_{\eta_0} \frac{d\eta_1}{\eta_1}\,\int^{\eta_1}_{\eta_0} \frac{d\eta_2}{\eta_2}\,\Bigg\{ J^+_{\vk}(\eta_1)J^+_{-\vk}(\eta_2)\,u(k,\eta_1)\,u^*(k,\eta_2)   \nonumber \\ & + &  J^-_{\vk}(\eta_1)J^-_{-\vk}(\eta_2)\,u^*(k,\eta_1)\,u(k,\eta_2)   -     J^+_{\vk}(\eta_1)J^-_{-\vk}(\eta_2)\,u^*(k,\eta_1)\,u(k,\eta_2) \nonumber \\ & - &   J^-_{\vk}(\eta_1)J^+_{-\vk}(\eta_2)\,u(k,\eta_1)\,u^*(k,\eta_2)\Bigg\}  \,.\label{Funfina2}\eea We seek to understand the emergence of a local effective action for super-Hubble wavelengths, in this case the product of mode functions of the environmental fields simplify to
\be u(k,\eta_1)\,u^*(k,\eta_2) \simeq \frac{1}{2\mu} \sqrt{\eta_1\eta_2}~ (\eta_1)^{i\mu} \,(\eta_2)^{-i\mu} \,.\label{produ}\ee Consider the following contribution for the first and last terms in (\ref{Funfina2})
\be \int^{\eta_1}_{\eta_*} d\eta_2 ~J^+_{-\vk}(\eta_2)\, \frac{\sqrt{\eta_1\eta_2}}{\eta_1\eta_2}(\eta_1)^{i\mu} \,(\eta_2)^{-i\mu} \label{terms} \ee where the lower limit $\eta_* \simeq -1/k$ so that only super Hubble fluctuations are considered. We generate an expansion in inverse powers of $\mu$ and derivatives by exploiting the identity
\be \frac{(\eta_2)^{-i\mu}}{\eta_2} = \frac{1}{i\mu}\,\frac{d}{d\eta_2} (\eta_2)^{-i\mu} \label{iden}\ee   integrating by parts  and neglecting the lower limit in the integrals. Implementing this systematically generates a series in $1/\mu$ and derivative terms.  In the long-time and super-Hubble limits the lower limit of the integrals at which the physical wavelength of the fluctuations is just of the order of the Hubble radius and the fluctuations did not yet amplify yields a subleading contribution. We find to leading and next to leading order that (\ref{terms}) becomes
\be J^+_{-\vk}(\eta_1)\Big[-\frac {1}{2i\mu^2}+\frac{1}{4\mu^3} \Big]+\frac{\eta_1}{2\mu^3} \,\frac{d}{d\eta_1}J^+_{-\vk}(\eta_1) +\cdots \label{leadplus}\ee  and the first term in the bracket in (\ref{Funfina2}) becomes
\be  \int^\eta_{\eta_*} d\eta_1 \, \Bigg\{ J^+_{\vk}(\eta_1)J^+_{-\vk}(\eta_1) \Big[-\frac {1}{2i\mu^2}+\frac{1}{4\mu^3} \Big]+\frac{\eta_1}{2\mu^3}J^+_{\vk}(\eta_1) \,\frac{d}{d\eta_1}J^+_{-\vk}(\eta_1)+\cdots \Bigg\}\,, \label{firsterm}\ee  a similar procedure yields for the second term
\be  \int^\eta_{\eta_*} d\eta_1 \, \Bigg\{ J^-_{\vk}(\eta_1)J^-_{-\vk}(\eta_1) \Big[ \frac {1}{2i\mu^2}+\frac{1}{4\mu^3} \Big]+\frac{\eta_1}{2\mu^3}J^-_{\vk}(\eta_1) \,\frac{d}{d\eta_1}J^-_{-\vk}(\eta_1)+\cdots \Bigg\}\,, \label{secterm}\ee to leading order we find that the third and fourth terms cancel each other out. We finally find that the influence action features both a real   and an imaginary (dissipative) local parts
\be \mathcal{F}[J^+, J^-] = \mathcal{F}_H[J^+, J^-]+\mathcal{F}_D[J^+, J^-] \label{realimagF}\ee given by
\be \mathcal{F}_H[J^+, J^-] = -\frac{g^2}{2\mu^2} \sum_{\vk} \int^\eta d\eta_1 \Bigg\{ J^+_{\vk}(\eta_1)J^+_{-\vk}(\eta_1)- J^-_{\vk}(\eta_1)J^-_{-\vk}(\eta_1) \Bigg\} \label{Fher}\ee
\be \mathcal{F}_D[J^+, J^-] = \frac{ig^2}{2\mu^3} \sum_{\vk} \int^\eta d\eta_1 \Big(J^+_{\vk}(\eta_1)-J^-_{\vk}(\eta_1)\Big)\Big(J^+_{-\vk}(\eta_1)-J^-_{-\vk}(\eta_1)\Big) \label{Fdiss}\ee where we have neglected the subleading derivative and imaginary terms in both expressions (suppressed by at least one power of $1/\mu  \ll 1$). Using systematically the identity (\ref{iden}) and a similar one for the $\eta_1$ terms one generates a series in powers of $1/\mu \ll 1 $, valid for $M_\varphi/H \gg 1$.

\begin{figure}[h!]
\includegraphics[height=4in,width=4in,keepaspectratio=true]{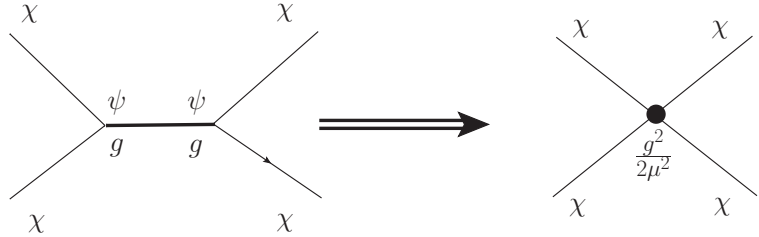}
\caption{Effective ``Fermi'' vertex in the case   for $J[\chi]=\chi^2~:~ \mathcal{O}[\psi]=\psi/\eta$ for $\mu = M_\varphi/H \gg 1$.}
\label{fig:effvertex}
\end{figure}

Comparison with (\ref{Seff}) clearly reveals that to leading order the   term (\ref{Fher}) simply yields a local effective action corresponding to a Lagrangian density
\be \mathcal{L}_{eff} =  \frac12\left[
{\chi'}^2-(\nabla \chi)^2-\mathcal{M}^2_{\chi}(\eta) \; \chi^2 \right]- \frac{g^2}{2\mu^2} \, \chi^4 \,. \label{Leffloc} \ee The effective quartic vertex in the local ``Fermi'' theory is shown in fig. (\ref{fig:effvertex}).

This local effective field theory is broadly in agreement with the results of refs.\cite{heavyachu,heavycespe,heavy1,heavy2,heavyjack}, however we emphasize that to next to leading order in $1/\mu$ there emerge \emph{dissipative} effects described by $\mathcal{F}_D$ in (\ref{Fdiss}) which are \emph{not} captured by the treatments in these references since this term emerges from the time evolution of the \emph{density matrix} and is not available from the simple path integral formulation in refs.\cite{heavyachu,heavycespe,heavy1,heavy2,heavyjack}.

\vspace{2mm}

\textbf{Interpretation:}

 \vspace{2mm}

The interpretation of the emergence of this local effective Lagrangian density becomes more clear recognizing the following aspects,
\begin{itemize}
\item with $\mu = \frac{M_\varphi}{H}$ and $g = \lambda/H $ (see eqn. (\ref{dimlessg}) with $p=1$) it follows that
    \be \frac{g^2}{\mu^2} = \frac{\lambda^2}{M^2_\varphi} \,, \label{effcoup}\ee
\item with $\eta = -e^{Ht}/H $ it follows that
\be (\eta_1)^{i\mu} \,(\eta_2)^{-i\mu} = e^{iM_\varphi(t_1-t_2)}\,,  \ee corresponding to the real time propagator of the heavy particle in the long wavelength limit.
\end{itemize}

In the limit $M_\varphi/H \gg 1$ the Compton wavelength of the heavy particle is much smaller than the Hubble radius and the long wavelength correlation functions of this field are the same as in Minkowski space time.

We emphasize that the emergence of a local ``Fermi'' limit is available \emph{only} when $M_\varphi/H \gg 1$, in the opposite limit $\mu \ll 1$  and implementing the steps described above would lead to a divergent series in powers of $1/\mu \gg 1$.

It is important to quantify what is the  mass scale that satisfies the assumption $M_\varphi \gg H$. \emph{If} the scale of the inflationary potential is the grand unified scale $\simeq (10^{16}\,GeV)^4$, it follows that $H \simeq 10^{14}\,GeV$ therefore $M_\varphi \gtrsim 10^{15}\,GeV$, thus we see that only degrees of freedom with masses of the order of or larger than the grand unified scale would fulfill this assumption. For degrees of freedom with masses well below $H$ which covers not only the standard model but most likely   physics beyond, the results for the cases with $M_\varphi \ll H$ are valid. Therefore, while the case of environmental fields with $M \gg H$ is of fundamental importance, on physical grounds it would be expected that there are many degrees of freedom with masses well below the grand unified scale that would yield stronger dissipative effects, unless for some (finely tuned) reason the inflaton is    \emph{not} directly coupled to these degrees of freedom.

\section{Discussion.}
Although in this study the large scale suppression of the power spectrum from dissipative effects have been obtained for the case of   a coupling of the form $\phi \varphi^2$ between the inflaton and a  massless   environmental field conformally coupled to gravity,  the results are likely of broader relevance. For example in ref.\cite{lee} a different coupling of the form $\phi^2\varphi^2$ was studied and the numerical results reported in this reference also suggest a suppression of the power spectrum. In ref.\cite{boyeff} such coupling is shown to lead to the case in which the noise is colored and multiplicative, a situation that requires a deeper analysis with the dynamical renormalization group. Furthermore, the modes of a conformally coupled massless fields represent faithfully the \emph{sub-Hubble limit} (indeed the high energy limit)  of mode functions for massive environmental scalar fields minimally coupled to gravity, irrespective of whether their mass is larger or smaller than $H$. Therefore the result obtained from the massless conformally coupled degrees of freedom is at the very least an important contribution to the self energy and noise correlator as is clear from the analysis of  minimally coupled fields that yield (\ref{totsig},\ref{totnoiss}) and which are further enhanced by infrared and secular divergences.

Although we discussed nearly massless scalar fields we expect the results to be broadly generalized to the case of fermionic environmental fields. In particular for  fermionic fields with masses $\ll H$ the mode functions are similar to the case of conformally coupled scalar fields\cite{uzan,boydVS,fermionswoodpro}, these mode functions are not amplified as the physical wavelength becomes super-Hubble. This is a manifestation of the fact that fermionic degrees of freedom are never classical. A Yukawa coupling between the inflaton fluctuations and fermionic fields yields an influence action in terms of self-energy and noise kernels much in the same manner as for the case studied in the previous sections. Results of this study will be reported elsewhere\cite{boynext}.

 Furthermore, our study focused on the power spectrum of the inflaton fluctuations, whereas it is the power spectrum of the curvature perturbation that is relevant for CMB anisotropies. In ref.\cite{kahya} the authors have studied loop corrections to the correlation function of the curvature perturbation from both   self-interaction   as well as the interaction with ``spectator'' scalar fields. In both cases the authors find logarithmic infrared corrections. From the results of ref.\cite{kahya}  considering, for example, a ``spectator'' potential $U(\phi) = \lambda \phi^2$ one infers a coupling $\zeta \phi^2$ with $\zeta$ proportional to the curvature perturbation\cite{kahya}, in which case the results obtained in the previous section are directly applicable to the case of curvature perturbations. Extrapolating our results to this particular example we would be led to conclude that the power spectrum for curvature perturbations is suppressed in the same manner as that of the inflaton fluctuations studied above. In a recent study\cite{tanaka}  the curvature perturbation is coupled to   heavy  scalar fields with $M \gg H$ minimally coupled to gravity  and the authors found that the power spectrum of curvature perturbations is \emph{not modified}. \emph{A prima facie} this result seems to contradict those in ref.\cite{kahya}, and also the fact that the contribution from sub-Hubble modes should be similar to the case of massless conformally coupled fields, as we found for minimally coupled fields. However the authors of ref.\cite{tanaka} enumerate several caveats that may invalidate their conclusions, chief among them is the violation of scale invariance. However, the breakdown of scale invariance  is precisely one of the hallmarks of our findings: for conformally coupled environmental fields this breakdown is a result of (powers of)  logarithmic secular terms consistently with Weinberg's theorem\cite{weinberg}, for minimally coupled fields is a consequence of both logarithmic secular terms \emph{and} logarithmic infrared singularities\cite{kahya,woodardcosmo,boyan}. Therefore the question of whether the power spectrum of curvature perturbations is suppressed from coupling to environmental degrees of freedom remains to be explored further and the results of ref.\cite{kahya} when combined with our analysis suggest  that the power spectrum of curvature perturbation will be suppressed by the coupling to environmental fields.

\section{Summary, conclusions and further questions}

The main question addressed in this article is what is the influence of sub-Hubble degrees of freedom that couple to the  inflaton  on   the  power spectrum of super-Hubble inflaton fluctuations. This question is motivated by puzzling large scale anomalies in the temperature power spectrum of the CMB. In a previous study\cite{boydensmat} the power spectrum of inflaton fluctuations were studied from the quantum master equation which describes the evolution of the reduced density matrix obtained from tracing out ``environmental'' degrees of freedom. The results of that study revealed a suppression of the power spectrum at large scales. In this article we study the problem from a different but complementary perspective by obtaining the non-equilibrium effective action for super Hubble inflaton fluctuations after integrating out (tracing out) ``environmental'' degrees of freedom coupled to the inflaton. The effective action includes the \emph{influence action} determined by the correlation functions of  the environmental fields and is obtained up to second order in the coupling for various relevant cases. This effective action is stochastic and yields a Langevin equation  of motion for the inflaton fluctuations with non-local self energy corrections and a noise term which explicitly describes the stochastic aspect of the dynamics induced by the coupling to environmental degrees of freedom.  For a coupling $\phi \varphi^2$ of the inflaton field $\phi$ to a scalar environmental field $\varphi$ the noise is additive and Gaussian but colored (non-local correlations) and the self-energy and noise correlation function obey a de Sitter space time generalization    of the fluctuation dissipation relation\cite{boyeff}. We model sub-Hubble environmental degrees of freedom by taking the environmental scalar field $\varphi$ to be massless and conformally coupled to gravity and solve the Langevin equation in the long time and super-Hubble limit for the inflaton fluctuations by implementing a dynamical renormalization group resummation of secular terms. We find the corrections to the power spectrum, assuming a (nearly) massless inflaton field
\be \mathcal{P}(k,\eta) = \Big(\frac{H}{2\pi}\Big)^2 e^{-\gamma(\eta)}~(1+ \mathcal{O}(g^2))   ~~;~~\gamma(k;\eta) = \frac{g^2}{12\pi^2}\Big[\ln^2\big( {-k\eta}  \big) - 2 \ln\big( {-k\eta} \big)\,\ln\big( {-k\eta_0}\big)  \Big]\,,\nonumber \ee where $g$ is the dimensionless coupling between the inflaton and the environmental fields. The dynamical renormalization group provides a non-perturbative resummation of Sudakov-type double logarithms leading to a suppression of the power spectrum via dissipative effects \emph{and a violation of scaling}.  The case of  environmental fields that are minimally coupled to gravity with $M_\varphi \ll H$ yields  stronger secular divergences as well as an infrared enhancement as a consequence of environmental degrees of super-Hubble wavelengths in the self-energy loop that are amplified. This case would lead to a stronger suppression of the power spectrum but possibly to contributions from isocurvature or entropy which are severely constrained by CMB observations.
We have also analyzed a coupling of the form $\phi^2 \varphi$ in the case in which the mass of the environmental field $M_\varphi \gg H$. In this case the long-wavelength effective action for the inflaton features a local ``Fermi limit'' with a very weak self-interaction $\phi^4$ and dissipative contributions which are suppressed by powers of $H/M_{\varphi}$. A simple interpretation of this Fermi limit result is given.

Our results confirm those obtained in ref.\cite{boydensmat} via the quantum master equation that describes the dynamics of the reduced density matrix, but complement those by providing directly a non-equilibrium effective action that leads to a stochastic description of the dynamics with precise derivation of the effective Langevin equation and the correlation functions of the stochastic noise, both completely determined by the correlation functions of the environmental fields.

There are both fundamental and practical corollaries of the study presented here:

\begin{itemize}
\item{ On the fundamental level, by reproducing the results of ref.\cite{boydensmat} we established a direct correspondence between the effective action including the environmental influence action and the quantum master equation that yields the dynamics of the reduced density matrix, and  established a direct relation with a stochastic description. An important result from the comparison between the two approaches is that the solution of the quantum master equation for correlation functions obtained from the reduced density matrix is equivalent to a dynamical renormalization group resummation.
    Both the effective action with the influence action and the quantum master equation provide complementary and equivalent powerful frameworks to study inflationary dynamics upon coarse graining or tracing out (sub-Hubble) degrees of freedom with which the inflaton field interacts. The influence action approach offers the advantage of coupling sources and obtaining correlation functions at different times, whereas the quantum master equation yields a direct approach to the equations of motion for expectation values in the reduced density matrix.  }

\item{ On the practical level: although we focused on the effective action of an inflaton-like scalar field in interaction with other scalars that are traced over (integrated out) leading to the \emph{influence action}, we \emph{conjecture} that the suppression of the power spectrum from dissipative effects associated with the interaction of sub-Hubble degrees of freedom is likely a robust feature that \emph{perhaps} may explain the puzzling large scale suppression of the CMB temperature power spectrum. The effective action with the influence action and the equivalent and complementary quantum master equation for the reduced density matrix provide a systematic non-perturbative framework to study the effects of ``environmental'' degrees of freedom on inflationary observables. }

\end{itemize}

 Important questions remain to be studied, among them the influence of different environmental degrees of freedom, such as fermions and the extension of the program presented here to curvature perturbations, not just to scalar inflaton fluctuations. Work on both fronts is in progress and will be reported elsewhere\cite{boynext}.

\acknowledgements  The author thanks D. Jasnow for enlightening discussions and  gratefully acknowledges the N.S.F. for partial
support through grants PHY-1202227, PHY-1506912.



\end{document}